\documentclass[aps,prl,twocolumn,superscriptaddress,groupedaddress]{revtex4-1}

\usepackage{bm}
\usepackage[usenames,dvipsnames]{color}
\usepackage{dcolumn}
\usepackage[usenames,dvipsnames,svgnames,table]{xcolor}
\usepackage{amssymb}
\usepackage{amsmath}
\usepackage{mathtools}
\usepackage{bbold}
\usepackage{amsfonts}
\usepackage{graphicx}
\usepackage{xstring}
\usepackage{units}
\usepackage{hyphenat}
\newcommand{\dt}{\frac{d}{dt}}
\newcommand{\tauon}{\tau_{\mathrm{on}}}
\newcommand{\D}{\mathit{\Delta}}
\renewcommand{\v}[1]{\boldsymbol{#1}}
\newcommand{\wo}[1]{_{\backslash #1}}
\newcommand{\T}{^\mathsf{T}}
\newcommand{\cp}[3][]{p\IfStrEqCase{#1}{{}{}}[#1](#2\,|\,#3)}
\newcommand{\normal}[2][]{\mathcal N(#2 \IfStrEqCase{#1}{{}{}}[;\,#1])}
\newcommand{\expect}[2][]{\left\langle \, #2\, \right\rangle_{#1} }
\newcommand{\var}[1]{\mathrm{Var}\left[#1\right]}
\newcommand{\tauref}{\tau_\mathrm{ref}}
\newcommand{\uthr}{\vartheta}  % {u_\mathrm{thr}}
\newcommand{\ueff}{u_\mathrm{eff}}
\newcommand{\taueff}{\tau_\mathrm{eff}}
\newcommand{\Erev}{E^\mathrm{rev}}
\newcommand{\El}{E_\mathrm{l}}
\newcommand{\gl}{g_\mathrm{l}}
\newcommand{\Cm}{C_\mathrm{m}}
\newcommand{\taubk}{\overline{\tau^\mathrm{b}_k}}
\newcommand{\tausyn}{\tau_\mathrm{syn}}
\newcommand{\gsyn}{g^\mathrm{syn}}
\newcommand{\gtot}{g_\mathrm{tot}}
\newcommand{\Isyn}{I^\mathrm{syn}}
\newcommand{\Jsyn}{J^\mathrm{syn}}
\newcommand{\Irec}{I^\mathrm{rec}}
\newcommand{\Iext}{I^\mathrm{ext}}
\newcommand{\Inoise}{I^\mathrm{noise}}
\newcommand{\ureset}{\varrho} %{u_\mathrm{reset}}
\newcommand{\erf}{\mathrm{erf}}
\newcommand{\thetaeff}{\vartheta_\mathrm{eff}}
\newcommand{\DKL}{D_\mathrm{KL}}
\newcommand{\Dkl}[2]{\textnormal{D}_\textnormal{KL}\left( #1 \,||\, #2 \right)}
\begin{document}

\preprint{APS/123-QED}

\title{Stochastic inference with spiking neurons in the high-conductance state}
\thanks{
The first two authors have contributed equally to this work.
We thank W.~Maass for his essential support, as well as S.~Habenschuss, M.~Brixner and the three anonymous reviewers for their helpful comments.
This research was supported by EU grants \#269921 (BrainScaleS), \#237955 (FACETS-ITN), \#604102 (Human Brain Project), the Austrian Science Fund FWF \#I753-N23 (PNEUMA) and the Manfred St\"ark Foundation.
}

\author{Mihai A. Petrovici$^1$*, Johannes Bill$^{1,2}$*, Ilja Bytschok$^1$, Johannes Schemmel$^1$, Karlheinz Meier$^1$}
\affiliation{$^1$Kirchhoff Institute for Physics, University of Heidelberg \\ $^2$Institute for Theoretical Computer Science, Graz University of Technology}

\date{\today}

\begin{abstract}
    The highly variable dynamics of neocortical circuits observed \textit{in vivo} have been hypothesized to represent a signature of ongoing stochastic inference, but stand in apparent contrast to the deterministic response of neurons measured \textit{in vitro}.
    Based on a propagation of the membrane autocorrelation across spike bursts, we provide an analytical derivation of the neural activation function that holds for a large parameter space, including the high-conductance state.
    On this basis, we show how an ensemble of leaky integrate-and-fire neurons with conductance-based synapses embedded in a spiking environment can attain the correct firing statistics for sampling from a well-defined target distribution.
    For recurrent networks, we examine convergence towards stationarity in computer simulations and demonstrate sample-based Bayesian inference in a mixed graphical model.
    This points to a new computational role of high-conductance states and establishes a rigorous link between deterministic neuron models and functional stochastic dynamics on the network level.
\end{abstract}

\pacs{xxx-xxx}
\maketitle

\subsection{Introduction}

In responding to environmental stimuli, brains have to make predictions based on incomplete, noisy and ambiguous data.
The recent hypothesis that the brain copes with this challenge by performing Bayesian, rather than logical inference \cite{kording2004bayesian, fiser2010statistically, friston2011action}, has been strengthened by electrophysiological data which identified neural correlates of the involved computations \cite{yang2007probabilistic,berkes2011spontaneous} and theoretical work on potential spiking network implementations \cite{deneve2008bayesian, buesing2011neural, rao2004hierarchical}.

In probabilistic inference, possible values of a quantity are described by a random \mbox{variable (RV) $z_k$}
and all dependencies between RVs are stored in a joint distribution $p(z_1,\dots,z_K)$.
The belief about a set of unobserved RVs $\{z_1, \dots, z_M\}$ given an observed set of RVs is represented by the posterior distribution $\cp{z_1,\dots,z_M}{z_{M+1},\dots,z_K}$.
In particular, the posterior contains information on the most likely conclusion, as well as on all potential alternatives.

With regard to the representation of probability distributions in the brain, theoretical work \cite{fiser2010statistically} has argued in favor of sample-based codes, i.e., instead of providing the entire distribution at any point in time, samples \mbox{$\v z^{(t)} \sim p(z_1,\dots,z_K)$} are used as a proxy.
When modeling large systems, this offers three important advantages.
First, approximate solutions can be provided at any time, with increasingly reliable results as the calculation progresses
(``anytime computing'').
Secondly, marginalization comes at no cost, as $p(z_k)$ can be determined by simply neglecting the values of all other RVs.
Thirdly, some sampling algorithms support a high degree of parallelization with an algorithmic structure that is reminiscent of neural %
networks \cite{hoyer2003interpreting}.
\begin{figure}[tbp]
    \centering
    \includegraphics[]{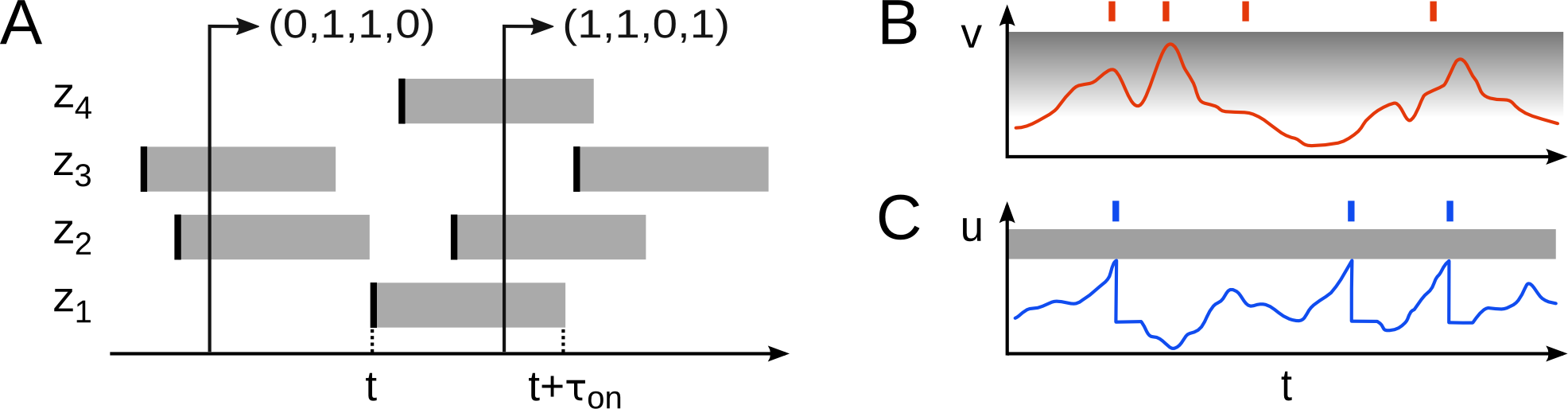}
    \caption{
        (A) Spike patterns as samples of a random \mbox{vector $\v z$}.
        The RV $z_k$ is active for duration $\tauon$ (gray bar) after each spike.
        (B) In stochastic neuron models, internal state variables (red) modulate the firing probability (gray).
        (C) In contrast, deterministic neurons elicit a spike when the membrane potential (blue) reaches a threshold voltage.
        \label{fig:1}
    }
\end{figure}
Recently, a theory has been suggested which combines these advantages by implementing Markov chain Monte Carlo (MCMC) sampling in spiking networks of abstract model \mbox{neurons \cite{buesing2011neural}}.
In this framework, spike patterns are interpreted as samples of binary RVs as follows (see Fig.~\ref{fig:1}A):
\begin{equation}
  z_k^{(t)} = 1 \Leftrightarrow \mathrm{Neuron} \; k \; \mathrm{fired \; in} \; (t-\tauon, t] \quad .
  \label{eqn:spikestostates}
\end{equation}
The duration $\tauon$ of the active state following a spike is a free parameter for the timescale on which a neuron affects downstream cells.
The neuron model in \cite{buesing2011neural} is inherently stochastic, with an instantaneous firing probability
\begin{align}
    \!\!r_k(t) \!=\!\! \lim_{\D t \to 0} \textstyle \frac{p(\mathrm{spike \; in \;} [t,\,t+\D t))}{\D t}
    = \left\{
    \begin{array}{ll}
        \frac{\exp(v_k)}{\tau} & \mathrm{if} \; z_k = 0 \\
        0 \quad\;\; & \mathrm{if} \; z_k=1
    \end{array}
    \right.
  \label{eqn:stochastic_neurons}
\end{align}
where $v_k$ is an abstract membrane potential (Fig.~\ref{fig:1}B).

In contrast to this idealized stochastic neuron model, \textit{in vitro} experiments have demonstrated the largely deterministic nature of single neurons \cite{mainen1995reliability}.
Similarly, microscopic models of neural circuits typically rely on deterministic dynamics of their constituents (see Fig.~\ref{fig:1}C).
The aim of this letter is to demonstrate how a network of deterministic neurons in a biologically plausible spiking environment can quantitatively reproduce the stochastic dynamics required for sampling from a well-defined distribution and perform inference given observations.
In this letter, we extend our earlier discussion of this approach \cite{petrovici2013stochastic}.

We start by calculating the dynamics of a single leaky integrate-and-fire (LIF) neuron in a spiking noisy environment and derive its activation function in the high-conductance state.
For this biologically relevant regime, existing analytical descriptions of neuronal response functions (e.g., \cite{brunel1998firingfrequency,moreno2004role}) are not applicable.
Here, we provide an approach based on the propagation of the membrane autocorrelation throughout spike bursts.
This establishes an equivalence to the abstract, inherently stochastic units (Eq.~\ref{eqn:stochastic_neurons}).
On the network level, we show how conductance-based synapses approximate the interaction for sampling from a well-defined target distribution.
Furthermore, we show that the distribution sampled by LIF networks remains a good approximation of the target distribution even for large networks with strong recurrent interaction.
We complement our study with a demonstration of probabilistic inference by implementing a small graphical model for pattern recognition in a recurrent network of LIF neurons.

\subsection{LIF dynamics in a high-conductance state}

We consider deterministic LIF neurons defined by \mbox{$\Cm\, \dt u_k = \gl (\El - u_k) + I_k$}, with membrane potential~$u_k$, capacitance~$\Cm$, leak potential~$\El$, leak conductance~$\gl$ and input current~$I_k$.
When $u_k$ crosses a threshold $\uthr$ from below, a spike is emitted and $u_k$ is reset to $\ureset$ for a refractory \mbox{period $\tauref$}.
We formally partition the total input current $I_k$ into recurrent synaptic input, diffuse synaptic noise and additional external currents: \mbox{$I_k = \Irec_k + \Inoise_k + \Iext_k$}.
The currents $\Irec_k$ and $\Inoise_k$ are mediated through synapses and obey $\Isyn_k = \sum_{i} g_{ki} \,(\Erev_i - u_k)$ with reversal potential $\Erev_i$ of the $i$th synapse.
The temporal evolution of the conductance $g_{ki}$ is modelled as a low-pass filter on pre-synaptic spikes: $\dt g_{ki}= - g_{ki} / \tausyn + \sum_s w_{ki} \, \delta(t-t_i^s)$, with synaptic time constant  $\tausyn$, weight $w_{ki}$ and spike times $t_i^s$.

We start by considering a single neuron that receives diffuse synaptic noise $\Inoise_k$ in the form of Poisson spike trains from its surrounding.
The capacity of recurrent networks to produce such noise has been shown in, e.g., \cite{destexhe2009self}.
For the following analysis of individual neurons in a noisy environment, we omit the index $k$ and set $\Irec = 0$.

When a conductance-based LIF neuron receives strong synaptic stimulation, it enters a so-called high-conductance state (HCS, \cite{destexhe03hcs}), characterized by accelerated membrane dynamics.
For an analytical treatment, it is advantageous to rewrite the membrane dynamics as \mbox{$\taueff \dt u = \ueff(t) - u$}, such that $u$ decays towards an effective leak potential $\ueff$ with an effective time constant \mbox{$\taueff = \Cm/\gtot$} \cite{richardson2005synaptic} (see Appendix AI).
The total conductance $\gtot$ subsumes both leakage and synaptic conductances.
In a high input rate regime, $\taueff \to 0$ and the effective \mbox{potential $\ueff$} simply becomes a linear transformation of the synaptic noise input (see AII).
Using methods similar to \cite{ricciardi1979ouprocess}, it can be shown that, in this regime, $\ueff(t)$ can be described as an Ornstein-Uhlenbeck (OU) process $du=\frac{1}{\tausyn}(\bar u - u)dt + \sigma dW$ with parameters (see AIII):
\begin{align}
    \bar u &= \textstyle \left\{\Iext + \gl\El + \sum_i \nu_i w_i \Erev_i \tausyn\right\} / \expect{\gtot} \quad , \label{eqn:oumu} \\
    \sigma^2 &= \textstyle \left\{\sum_i \nu_i \left[ w_i \left( \Erev_i - \bar u \right) \right]^2 \tausyn\right\} / \expect{\gtot}^2 \quad . \label{eqn:ousigma}
\end{align}

\subsection{The activation function of LIF neurons in the HCS}

\begin{figure}[t]
    \centering
    \includegraphics[]{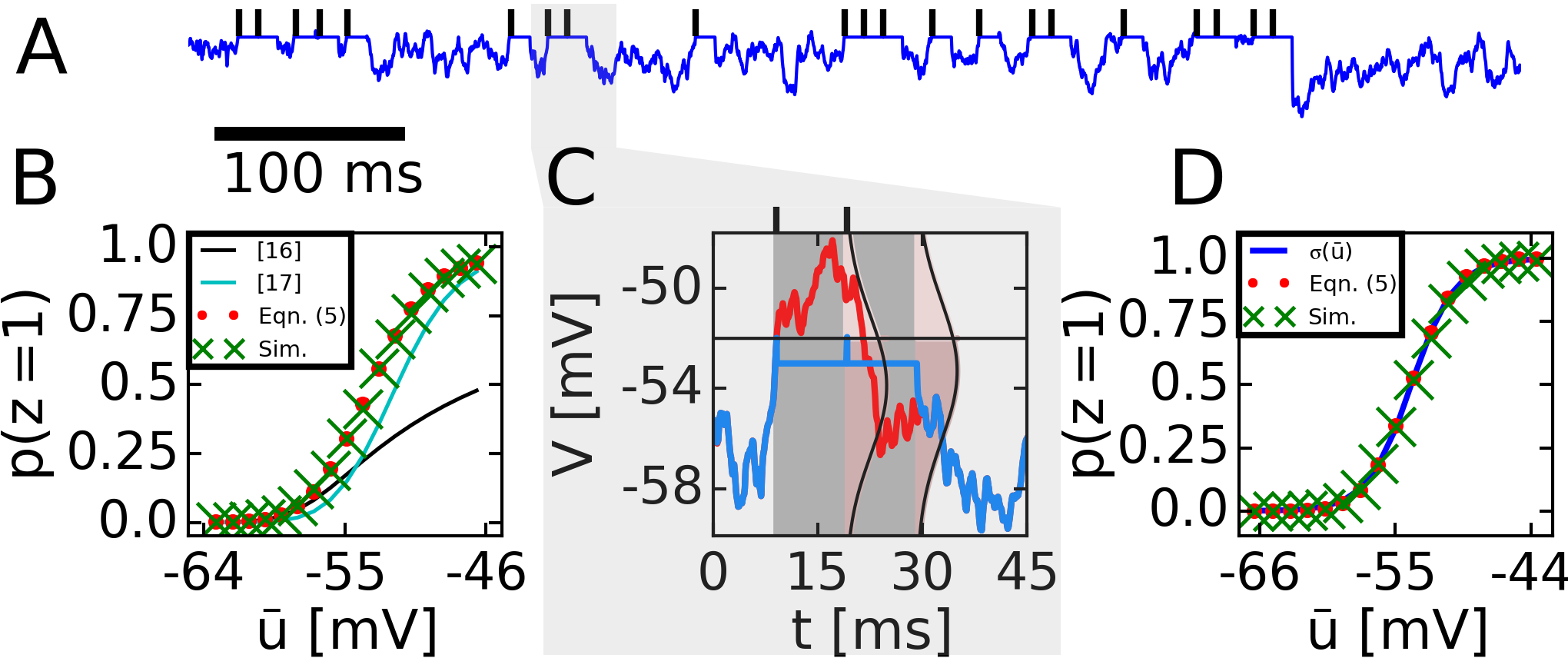}
    \caption{
        (A) Membrane potential $u(t)$ and spikes of an LIF neuron in a spiking noisy environment.
        (B) Prediction of the activation function (red) compared to simulation results (green), as well as to other predictions from literature \cite{brunel1998firingfrequency,moreno2004role}.
        (C) In a HCS, $u$ (blue) and $\ueff$ (red) are nearly identical when the neuron is not refractory.
        After each refractory period (gray), the predicted distribution of $\ueff$ (pink) is used for the propagation \mbox{in (Eq.~\ref{eqn:fullrecursion1_maintext})}.
        (D) High-noise regime: theoretical prediction (red) vs. simulation results (green), with a fitted logistic function $\sigma(\bar u)$ (blue).
    }
    \label{fig:2}
\end{figure}

Similarly to the abstract model \cite{buesing2011neural}, we define the refractory state of a neuron as $z(t) = 1$.
The mapping of spikes to RV states (Eq.~\ref{eqn:spikestostates}) naturally leads to the concept of an activation function $p(z=1|\bar{u})$, where $\bar u = \expect[t]{\ueff}$.
In the following, we derive a general expression for the activation function of an LIF neuron under Poisson stimulus, which is of particular use in the cortically relevant HCS regime.
Fig.~\ref{fig:2}A shows an exemplary simulation (membrane potential and spike train) of such a scenario.
The activation function in Fig.~\ref{fig:2}B is obtained by sweeping over $\Iext$ (see AVII and AVIII).

Related setups have already been examined in \cite{brunel1998firingfrequency,moreno2004role}.
However, the methods in \cite{brunel1998firingfrequency,moreno2004role} are tailored to certain parameter ranges, which, in particular, do not include the HCS regime with refractoriness.
This is because \cite{brunel1998firingfrequency} requires $\tausyn \ll \taueff$, whereas \cite{moreno2004role} assumes $\tauref \ll \taueff, \tausyn$, which leads to discrepancies in the predicted activation functions (Fig.~\ref{fig:2}B).
The deeper reason for the observed discrepancies is found in the lack of an appropriate propagation of the autocorrelation of $\ueff$ through $\tauref$.
Here, we propose a derivation that explicitly includes this propagation and thereby covers a large parameter space, including the cases studied in \cite{brunel1998firingfrequency,moreno2004role}, as well as the HCS.

In Fig.~\ref{fig:2}A, two modes of firing can be observed: a ``bursting'' mode, where the effective membrane potential $\ueff$ after the refractory period is still above threshold, and a freely evolving mode, where the neuron does not spike again immediately after the refractory period.
This is illustrated by the distributions in Fig.~\ref{fig:2}C.
Our approach relies on the calculation of burst lengths and their associated occurrence probability $P_n$.

Denoting the relative occurrence of burst lengths $n$ by $P_n$, the average drift time from $\ureset$ to $\uthr$ between the $k$th and $(k+1)$st spike in a burst by $\taubk$ and the average duration of the freely evolving mode that follows an $n$-spike-burst by $T_n$, we identify the following relation:
\begin{equation}
    p(z = 1|\bar u) = \frac{\sum_n P_n \cdot n \cdot \tauref}{\sum_n P_n \cdot \left(n\tauref + \sum_{k=1}^{n-1} \taubk + T_n\right)} \quad .
    \label{eqn:activationburstsum_maintext}
\end{equation}
Knowing the OU process governing $\ueff$, recursive expressions for $P_n$ and $T_n$ can be derived.
These terms have considerable impact on the activation function if $\tauref \approx \tausyn$.
In the limit of strong noise stimuli, we can calculate the average drift time $\taubk$ in a quasistatic approximation \cite{moreno2004role} of $\ueff$ and assume that $u \approx \ueff$ for the freely evolving mode, thus obtaining (see AIV):
\begin{align}
    P_n     & = \textstyle \left( 1 - \sum_{i=1}^{n-1} P_i \right) \int_{\uthr}^\infty du_{n-1} p(u_{n-1} | u_{n-1} \geq \uthr) \nonumber \\
            & \textstyle \quad \times \left[ \int_{-\infty}^{\uthr} du_n p(u_n | u_{n-1}) \right] \; , \label{eqn:fullrecursion1_maintext} \\
    T_n     & = \textstyle \int_{\uthr}^\infty du_{n-1} p(u_{n-1} | u_{n-1} \geq \uthr) \nonumber \\
            & \textstyle \quad \times \left[ \int_{-\infty}^{\uthr} du_n p(u_n | u_n < \uthr, u_{n-1}) \expect{T(\uthr, u_n)} \right] \; , \label{eqn:fullrecursion2_maintext} \\
    \taubk  &= \textstyle \int_\uthr^\infty du_k \ln \left(\frac{\ureset-u_k}{\uthr - u_k}\right) \nonumber \\
            & \textstyle \quad \times \int_\uthr^\infty du_{k-1} p(u_k | u_k \geq \uthr, u_{k-1}) \; .
            \label{eqn:fullrecursion3_maintext}
\end{align}
Fig.~\ref{fig:2}C displays an intuitive picture of the integrals in (Eq.~\ref{eqn:fullrecursion1_maintext}) and (Eq.~\ref{eqn:fullrecursion2_maintext}).
The transfer function $p(u_n | u_{n-1})$ is the Green's function of the OU process at time $t-t_s=\tauon$ and $\expect{T(\uthr, u_n)}$ denotes the average time needed for the membrane to reach $\uthr$ starting from $u_n$, which can be given in closed form \cite{ricciardi1988fptdensity}.
The dependency of $P_n$, $T_n$ and $\taubk$ on the moments of the OU process (Eq.~\ref{eqn:oumu},\ref{eqn:ousigma}) renders Eq. (\ref{eqn:activationburstsum_maintext}) a function of $\bar u$.

So far, we have only considered $\taueff=0$.
To further improve the prediction, we take into account finite values of $\taueff$ by means of an expansion in $\sqrt{\taueff/\tausyn}$.
Due to the symmetry of the PSP shape in $\taueff$ and $\tausyn$, this can be done analogously to \cite{brunel1998firingfrequency,moreno2004role}, where the opposite limit of large $\taueff$ and small $\tausyn$ is used (see AIV).
A comparison between our prediction of $p(z=1|\bar u)$ and results from a numerical simulation is shown in Fig.~\ref{fig:2}B.

\subsection{Sampling with networks of LIF neurons}

\begin{figure}[!t]
    \centering
    \includegraphics[width=8.6cm]{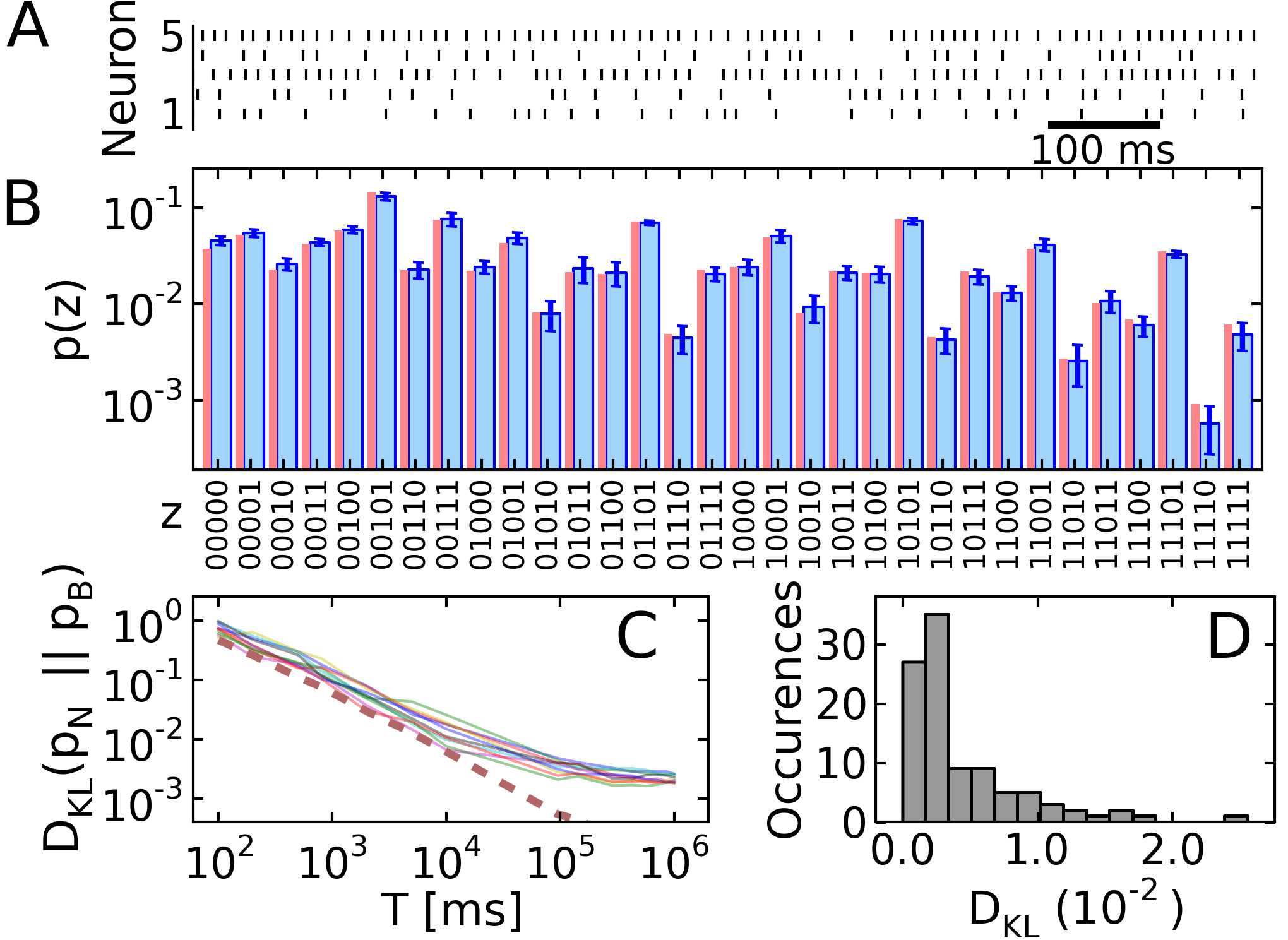}
    \caption{\label{fig:3}
        (A) Spike pattern of a recurrent network of $K=5$ LIF neurons during sampling from a randomly generated Boltzmann machine.
        (B) Sampled distribution $p_\mathrm{N}(\v z)$ of network states (blue) and target distribution $p_\mathrm{B}(\v z)$ (red).
        $p_\mathrm{N}(\v z)$ was estimated from ten \unit[10]{s} simulation runs (errorbars: std.\ deviation between runs).
        (C) $\Dkl{p_\mathrm{N}}{p_\mathrm{B}}$ as a function of integration time $T$ for 10 trials.
        The red dotted line shows convergence for the theoretically optimal abstract model \cite{buesing2011neural}.
        (D) $\Dkl{p_\mathrm{N}}{p_\mathrm{B}}$ when sampling for $T=\unit[10^6]{ms}$ from 100 different randomly generated target distributions.
      }
\end{figure}

We can now reconcile the response of LIF neurons with the inherently stochastic neuron model (Eq.~\ref{eqn:stochastic_neurons}) that requires a logistic activation function for constant potential~$v$: $p(z = 1|v) = \sigma(v) :=\left[ 1+\exp(-v) \right]^{-1}$.
In the HCS regime, this logistic activation can be approximated by LIF neurons with high accuracy.
Fig.~\ref{fig:2}D shows our theoretical prediction and simulation results for the activation function alongside a fitted logistic function.
For the translation from the LIF domain to the abstract model (Eq.~\ref{eqn:stochastic_neurons}) we have employed a linear mapping
\begin{align}
    v &= (\bar u - \bar u^0) / \alpha \quad ,
    \label{eq:linear_trans_membrane}
\end{align}
with scaling factor $\alpha$ and $\bar u^0$ denoting the potential for which \mbox{$p(z=1) = \frac{1}{2}$}.

We next connect the neurons to form a recurrent network.
In addition to noise stimuli, an LIF neuron in a network receives synaptic currents $\Irec_k$ from other neurons.
For certain connectivity structures, it is possible to predict the target distribution \cite{buesing2011neural,pecevski2011probabilistic} of states $\v z^{(t)}$ that arise from the stochastic dynamics of the recurrent network.
\begin{figure*}[]
    \centering
    \includegraphics[width=\textwidth]{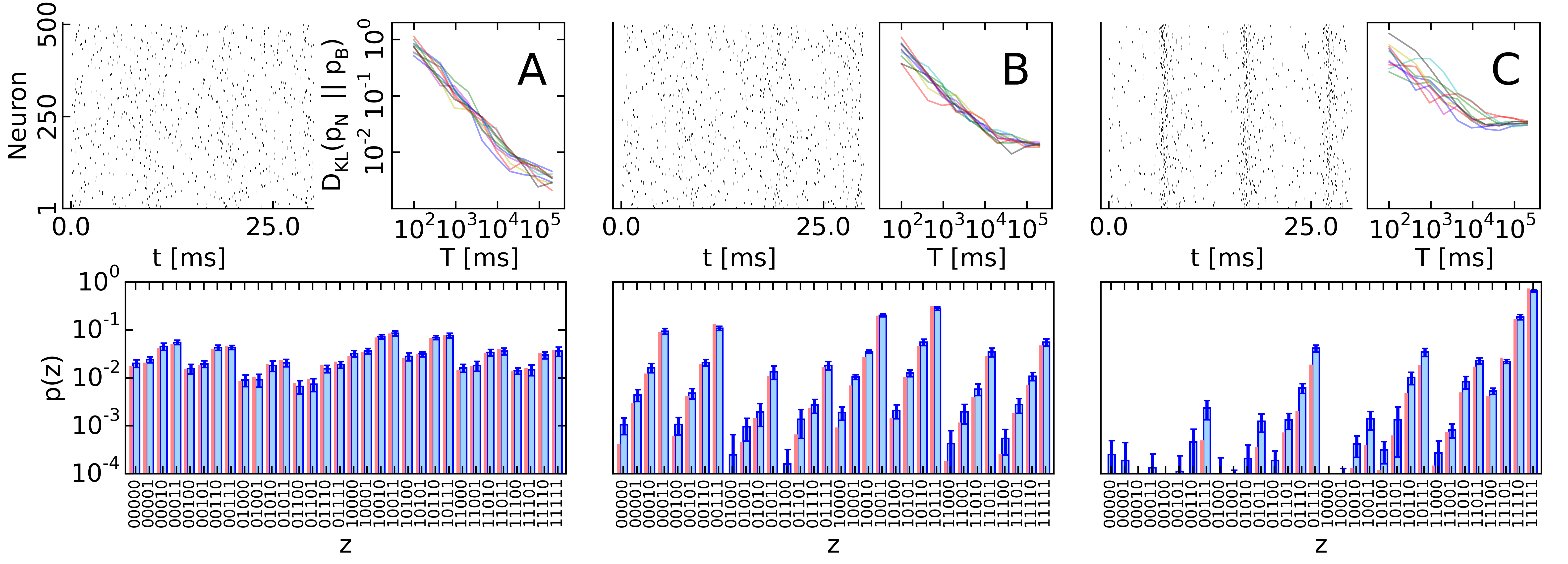}
    \caption{\label{fig:3large}
        Sampling in large networks.
        Spike raster (top left), joint distribution (bottom) and KL divergence (top right) as in Fig.~\ref{fig:3}A,B,C.
        Since the target distribution cannot be computed analytically for 500 RVs, we define the Gibbs-sampling estimate after $10^6$ steps as a reference distribution.
        For the joint, the sampled distribution over 5 RVs (out of 500) obtained from the LIF network (after $T_\mathrm{sim} = \unit[10^4]{ms}$) is plotted alongside the Gibbs estimate.
        Error bars indicate standard deviation between 10 simulation runs.
        In all panels, biases were drawn from a beta distribution: $b_k \sim 1.2 \cdot \left[ \mathcal B(0.5, 0.5) - 0.5 \right]$.
        (A) Small weights: $W_{kj} \sim 0.6 \cdot \left[ \mathcal B(0.5, 0.5) - 0.5 \right]$.
        (B) Moderate weights: $W_{kj} \sim 1.2 \cdot \left[ \mathcal B(0.5, 0.5) - 0.5 \right]$.
        (C) Strong weights: $W_{kj} \sim 2.4 \cdot \left[ \mathcal B(0.5, 0.5) - 0.5 \right]$.
        Systematic deviations between the sampled distribution and the reference manifest mainly in low-probability modes and are due to the difference in PSP shapes between the abstract model and LIF neurons.
            }
\end{figure*}
We use the emulation of Boltzmann machines (BMs) as an example case.
The joint distribution reads:
\begin{equation}
    \label{eq:BM_joint}
    p_\mathrm{B}(\v z) = Z^{-1}\,\exp\left(\v z\T \v W \v z /2+ \v z\T \v b \right) \quad ,
\end{equation}
where $\v W$ is a symmetric zero-diagonal weight matrix, $\v b$ is a bias vector and $Z$ is the normalizing partition function.
This probabilistic model underlies state-of-the-art machine learning algorithms for image \cite{salakhutdinov2009deep} and speech recognition \cite{mohamed2012acoustic}.
It has been shown \cite{buesing2011neural} that a network of abstract neurons (Eq.~\ref{eqn:stochastic_neurons}) with linear membrane potentials
\begin{equation}
    \label{eq:BM_vmem}
    v_k = b_k + \textstyle \sum_{j=1}^K W_{kj}\,z_j
\end{equation}
will sample from the desired target distribution (Eq.~\ref{eq:BM_joint}).
This finding uses the fact that individual neurons sample from the conditionals \mbox{$\cp{z_k = 1}{\v z \wo k} = \sigma(v_k)$}, with \mbox{$\v z \wo k = \{ z_j \, | \, j \neq k \}$}, in an MCMC updating scheme.

As shown above, LIF neurons in a spiking noisy environment closely approximate this logistic activation function if the synaptic currents $\Irec_k$ shift the mean membrane potential $\bar u_k$ according to the linear interaction (Eq.~\ref{eq:BM_vmem}).
Using the linear transformation (Eq.~\ref{eq:linear_trans_membrane}) between $v_k$ and $\bar u_k$, and estimating the effect of a conductance-based synapse of weight $w_{kj}$, we arrive at the following translation between the abstract and the LIF domain (see AV):
\begin{align}
    b_k & = (\bar u^b_k - \bar u^0_k) / \alpha \label{eqn:biastrans}\\
    W_{kj} & = \frac{1}{\alpha \Cm} \frac{w_{kj} \left(\Erev_{kj} - \mu\right)}{\frac{1}{\tausyn} - \frac{1}{\taueff}} \nonumber \\
           & \quad \times \left[\frac{1-e}{e} \!-\! \frac{\taueff}{\tausyn} \left( e^{- \frac{\tausyn}{\taueff}} - 1 \right) \right] ,
    \label{eq:weight_trans}
\end{align}
where $\bar u^b_k$ is the mean potential $\bar u_k$ that establishes $p(z_k=1|\bar u_k=\bar u_k^b) = \sigma(b_k)$ in Eq.~(\ref{eqn:activationburstsum_maintext}), and $\Erev_{kj}$ denotes the reversal potential for synapse $w_{kj}$.
The idea behind (Eq.~\ref{eq:weight_trans}) is to match the integrals of individual postsynaptic potentials (PSPs) on $v_k$ and $\bar u_k$.
Since the membrane loses any memory following a reset, in contrast to \cite{buesing2011neural}, we use the synaptic conductance as a memory carrier.

The remaining systematic difference to the abstract model lies in the additive -- instead of renewing -- nature of PSPs elicited by the same presynaptic neuron, which has a noticeable effect in case of fast consecutive spikes (bursts).
Since the membrane potential closely follows the effective potential in the HCS, renewing PSPs can be achieved by using renewing postsynaptic conductances.
For this, we have used short-term synaptic depression \citep{tso97} with a recovery time constant equal to $\tausyn$.

\begin{figure*}[]
  \centering
  \includegraphics[width=\textwidth]{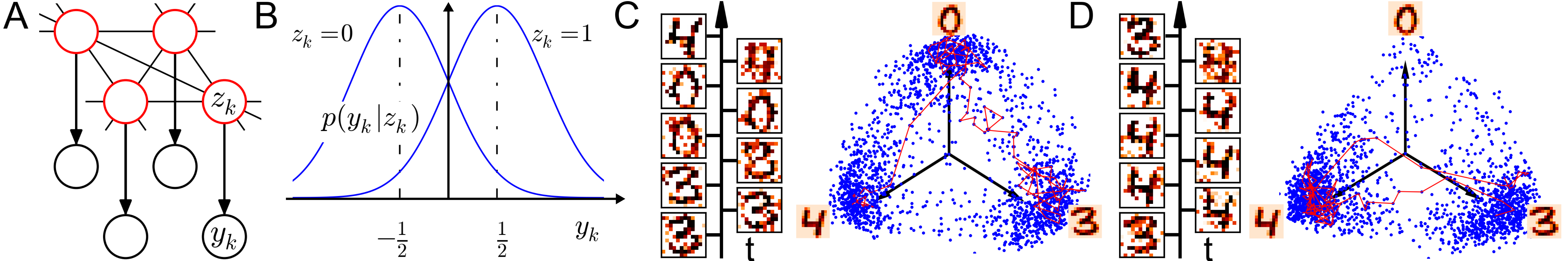}
  \caption{\label{fig:4}
      (A) Graphical model used for the probabilistic inference task.
      (B) A Gaussian likelihood model provides input to the sampling neurons.
      (C) Two dimensional projection of states $\v z^{(t)}$ when sampling from the prior $p(\v z)$, which was trained to store hand-written digits $0$, $3$ and $4$.
      Solid line: network trajectory over $\unit[200]{ms}$.
      Color maps: Marginals of $z_k$ averaged over $\unit[25]{ms}$.
      The time arrow covers the duration of the red trajectory and consecutive snapshots are $\unit[25]{ms}$ apart.
      (D) As in (C) when provided with incomplete input $\v y$ that is incompatible with digit $0$ and ambiguous with respect to digits $3$ and $4$.
      }
\end{figure*}

The sampling quality with networks of LIF neurons was examined in computer simulations of BMs, with randomly drawn parameters $b_k$ and $W_{kj}$ (Fig.~\ref{fig:3}A,B).
The target distribution $p_\mathrm{B}(\v z)$ is approximated by the distribution $p_N(\v z)$ of network states with $\tau_\mathrm{on} = \unit[10]{ms}$.
The chosen integration time $T=\unit[10]{s}$ displays a conservative estimate of the maximum duration a neuronal ensemble will experience stable stimulus conditions in a behaving organism and can thus be expected to sample from a stable target distribution.
For this integration time, we find that the recurrent network of LIF neurons accurately encodes the target distribution, within the precision imposed by the sample-based representation.
The network distribution $p_N(\v z)$ becomes increasingly more reliable as more samples are considered (Fig.~\ref{fig:3}C).
After few samples, the network has generated a coarse approximation of $p_\mathrm{B}(\v z)$ that could serve as an ``educated guess'' in online computation tasks.
Only for simulation times $T$ well beyond biologically relevant timescales do systematic errors in $p_N(\v z)$ become apparent.
The sampling quality holds for a variety of target distributions (Fig.~\ref{fig:3}D, see AIX for simulation details).

These observations hold for larger-scale networks as well (Fig.~\ref{fig:3large}).
For weak coupling, the networks are in an asynchronous irregular state of firing, which enables highly accurate sampling of the target distribution.
As synaptic weights increase, the network activity expectedly becomes more synchronous and the sampled distribution overall less accurate, especially in low-probability regions of the state space, but the high-probability modes are always sampled from with high fidelity.

\subsection{Demonstration of probabilistic inference}

We conclude our investigation of sampling in recurrent networks of LIF neurons with an example of Bayesian inference based on incomplete observations.
A fully connected BM of $K=144$ neurons, aligned on a $12\times12$ grid, was trained as an associative network \cite{Hinton26051995} to store patterns of hand-written digits $0$, $3$ and $4$ in the weights $W_{kj}$ and biases $b_k$.
Each pixel of the image grid was assigned to one network neuron.
The resulting joint distribution $p(\v z)$ displays ``prior knowledge'' stored by the network.

The probabilistic model was augmented by adding real-valued input channels for each pixel, associated with random variables $y_k \in {\mathbb R}, \, 1 \le k \le K$.
The resulting generative model \mbox{$p(\v y,\, \v z)$} has the structure shown in Fig.~\ref{fig:4}A and connects the latent network variables~$z_k$ to observable inputs $y_k$ by means of likelihood functions $\cp{y_k}{z_k}$, which we have chosen to be Gaussian with unit variance (Fig.~\ref{fig:4}B).
The likelihoods $\cp{y_k}{z_k}$ tend to align the network state with the observation, i.e.\ $z_k=1$ for $y_k > 0$,
while the prior $p(\v z)$ reconciles the observations with knowledge on consistent activation patterns~$\v z$.
The task for the network is to calculate and represent the posterior distribution according to Bayes' rule: \mbox{$\cp{\v z}{\v y} \propto p(\v z) \cdot \cp{\v y}{\v z}$}.
A short derivation (see AVI) shows that the posterior $\cp{\v z}{\v y}$ is a BM for any input $\v y$ with the following abstract membrane potential:
\begin{align}
  \textstyle v_k = b_k + y_k + \sum_{j} W_{kj} \, z_j \;\;.
\end{align}
In the LIF domain, the sum $b_k + y_k$ is equivalent to an effective bias (Eq.~\ref{eqn:biastrans}) and corresponds to an external current $\Iext_k = I^b_k + I^y_k$ that shifts $\bar u_k$ appropriately.
Thus, a neuron receives synaptic input from recurrent connections and noise sources, as well as an external current, i.e.,  $I_k = \Irec_k + \Inoise_k + \Iext_k$.

In case of $I^y_k = 0\;\forall k$, the network samples from the prior distribution $p(\v z) = \cp{\v z}{\v y = \v 0}$.
A two-dimensional projection of network states $\v z^{(t)} \sim p(\v z)$ is shown in Fig.~\ref{fig:4}C.
The sampled distribution has three distinct modes that correspond to the three hand-written digits stored in the recurrent weight matrix.
A closer look at the network trajectory reveals that the system stays in one mode (``digit'') for some duration, traverses the state space and then samples from a different mode of the distribution.

A typical inference scenario with incomplete observations is shown in Fig.~\ref{fig:4}D.
Four input channels at the center were picked to inject positive currents $I^y_k > 0$ to the network while all other inputs remained uninformative.
Positive currents $I^y_k$ were chosen such that the observation $\v y$ appeared incompatible with digit 0, and remained ambiguous with respect to digits 3 and 4 (see AX).
In accordance with Bayes' rule, the resulting bimodal posterior distribution $\cp{\v z}{\v y}$ has a suppressed $0$-mode, but preserves the $3$ and $4$ modes.

\subsection{Discussion}

We have shown how recurrent networks of conductance-based neurons in a spiking noisy environment can perform probabilistic inference through sampling from a well-defined posterior distribution.
Our approach extends Bayesian spiking network implementations to deterministic neuron models widely used in computational neuroscience.
We have provided an analytical derivation of the bursty firing response of LIF neurons under Poisson bombardment, which holds for a wide range of parameter regimes (both high and low ratios of $\tausyn/\taueff$, as well as both high and low $\tauref$).
Our approach is based on the propagation of the membrane autocorrelation throughout bursts and can thereby provide a prediction of the activation function in a regime where existing approaches \cite{brunel1998firingfrequency,moreno2004role} do not hold (see also Fig. \ref{fig:actfctcomp} in AXI).
We have further shown how high-frequency spiking inputs that could be provided by the surrounding network can lead to fast membrane dynamics, which enable individual LIF neurons to correctly encode conditional distributions given information from their presynaptic partners.
Thereby, our derivation also identifies a potential functional role of biologically observed high-conductance states and synaptic memory within a Bayesian framework of brain computation.

For mathematical tractability, simplifying modeling assumptions had to be made.
The neuron model uses an absolute refractory time $\tauref$, which matches the activation time constant $\tauon$, and neglects any gradual recovery effects.
On the network level, we have assumed statistically independent noise sources and instantaneous axonal transmission.
One important difference to cortical structure is the requirement of a symmetric connectivity matrix.
The precise symmetry is a consequence of all neurons sharing the same parameters and can be relaxed as neurons become diverse ($\alpha\mapsto\alpha_k$ in Eq.~\ref{eq:weight_trans}).
Furthermore, while Dale's principle is known to not hold universally \citep{sulzer2000dale}, negative coupling in large networks between otherwise excitatory neurons could be, in principle, introduced through populations of inhibitory interneurons.
LIF PSPs differ from the theoretically optimal rectangular shape, which could impair convergence to the target distribution outside of the high noise regime \cite{gerstner1992associative}.
However, computer simulations indicate that in many biologically relevant scenarios the above approximations are not critical (see also Fig.~\ref{fig:paramsweep} in AXII).
In particular, the sampling properties of LIF networks remain preserved as network size and interaction strength increase (Fig.~\ref{fig:3large}).
Embedding LIF sampling in cortical-size networks appears feasible in light of our results, but remains a matter for future work.

For neuroscientific modeling, our analysis of LIF neurons can be readily transferred to other neuron \cite{brette2005adaptive} and synapse models.
Beyond neuroscience, the ability to perform probabilistic inference with deterministic neurons displays a promising computing paradigm for neuromorphic hardware systems, which typically implement physical models of integrate-and-fire neurons \cite{mitra2009real,petrovici2014characterization}.
The distributed nature of the proposed LIF sampling networks allows to exploit the inherent parallelism of neuromorphic architectures and fosters their application to online data evaluation and robotics.
In this context, our results have already provided the basis for the implementation of Bayesian networks \cite{probst2015probabilistic} and learning \cite{neftci2014event-driven}.

\vfill

\section{APPENDIX}

\subsection{Appendix I:\\ Conductance-based LIF neuron model}

We restate the set of equations that govern the conductance-based LIF model.
For the membrane potential, we have
\begin{equation}
    \Cm\, \frac{du}{dt} = \gl (\El - u) + I \quad ,
    \label{eqn:LIF_dynamics}
\end{equation}
with capacitance $\Cm$, membrane potential $u_k$, leak potential $\El$, leak conductance $\gl$ and input current $I$.
The input current $I$ can be formally partitioned into recurrent synaptic inputs, diffuse synaptic noise and additional ``external'' currents, i.e.,
\begin{equation}
    I = \Irec + \Inoise + \Iext \quad .
\end{equation}
$\Iext$ may represent external current stimuli, average synaptic stimulus currents, or changes in the leak mechanism (i.e., changes in $\gl$ or $\El$).
The total synaptic current $\Isyn = \Irec + \Inoise$ obeys the equation
\begin{equation}
    \Isyn = \sum_{\mathrm{syn} \, i} \gsyn_i \left( \Erev_i - u \right) \quad ,
\end{equation}
where $\gsyn_i$ represents the conductance at the $i$th synapse and $\Erev_i$ the corresponding reversal potential.
The synaptic conductance obeys the ODE
\begin{equation}
    \frac{d\gsyn_i}{dt} = -\frac{\gsyn_i}{\tausyn} + \sum_{\mathrm{spk} \, s} w_i \delta(t-t_s) \quad ,
    \label{eqn:gsynODE}
\end{equation}
with synaptic time constant  $\tausyn$ and weight $w_{i}$.
The sum runs over all presynaptic spikes $s$.
The solution to this equation is a superposition of exponentials:
\begin{equation}
    \gsyn_i = \sum_{\mathrm{spk} \, s} w_i \Theta(t-t_s) \exp\left(-\frac{t-t_s}{\tausyn}\right) \quad .
    \label{eqn:g(t)}
\end{equation}
Putting all of the above together, we obtain the full ODE for the membrane potential
\begin{align}
    \Cm\, \frac{du}{dt} &= \gl (\El - u) \nonumber \\
                        &+ \sum_i \sum_{\mathrm{spk} \, s} w_i \Theta(t-t_s) \exp\left(-\frac{t-t_s}{\tausyn}\right) \left( \Erev_i - u \right) \nonumber \\
                        &+ \Iext \quad .
    \label{eqn:u_dynamics}
\end{align}
We can now divide the RHS of (Eq. \ref{eqn:u_dynamics}) by $\gtot = \gl + \sum_i \gsyn_i$ and rearrange the terms in order to obtain
\begin{equation}
    \taueff \frac{du}{dt} = \ueff - u \quad ,
    \label{eqn:reduceddudt}
\end{equation}
with a new effective membrane time constant
\begin{equation}
    \taueff = \frac{\Cm}{\gtot}
\end{equation}
and effective leak potential
\begin{equation}
    \ueff(t) = \frac{\gl\El + \sum_i \gsyn_i(t) \Erev_i + \Iext}{\gtot(t)} \quad .
    \label{eqn:ueff}
\end{equation}
This transformation is routinely used in studies of conductance-based neurons (see, e.g., \cite{richardson2005synaptic}).
Here, we have made the time dependencies explicit, since the following HCS approximation will serve to eliminate $t$ in the denominator.

\subsection{Appendix II:\\ The high-conductance state (HCS)}

In a first approximation, assuming a rapidly-firing Poisson background ($\nu_\mathrm{syn} \rightarrow \infty$), the total average conductance can become arbitrarily large ($\expect{\gtot} \rightarrow \infty$), causing the membrane potential to follow the effective potential nearly instantaneously ($\expect{\taueff} \rightarrow 0$). Eq. (\ref{eqn:ueff}) can then be rewritten as
\begin{equation}
    u \approx \ueff = \frac{\gl\El \! + \! \sum_i \! \expect{\gsyn_i} \! \Erev_i \! + \! \sum_i \Delta \gsyn_i \Erev_i \! + \! \Iext}{\expect{\gtot} + \sum_i \Delta \gsyn_i} ,
    \label{eq:ufromg}
\end{equation}
where
\begin{equation}
    \Delta \gsyn_i = \gsyn_i - \expect{\gsyn_i}
\end{equation}
denotes the fluctuations of the synaptic conductances.

For a single Poisson source with rate $\nu_i$ connected to the neuron by a synapse with weight $w_i$ and time constant $\tausyn$, the conductance course can be seen as a sum of independent random variables, each of them representing the conductance change caused by a single spike.
In the limit of large $\nu_i$, the central limit theorem guarantees the convergence of the conductance distribution to a Gaussian, with moments given by
\begin{align}
    \expect{\gsyn_i} &= \sum_{\mathrm{spk} \, s} \expect{ w_i \Theta(t-t_s) \exp \left(-\frac{t-t_s}{\tausyn} \right) } \nonumber \\
                     &= \lim_{T \to \infty} \frac{\expect{N}}{T} w_i \int_0^T \exp \left( -\frac{t}{\tausyn} \right) \, dt \nonumber \\
                     &= w_i \nu_i \tausyn
\end{align}
and
\begin{align}
    \var{\gsyn_i} &= \sum_{\mathrm{spk} \, s} \var{ w_i \Theta(t-t_s) \exp \left(-\frac{t-t_s}{\tausyn} \right) } \nonumber \\
                  &= \lim_{T \to \infty} \expect{N} \left\{ \expect{ \left[ w_i \Theta(t-t_s) \exp \left(-\frac{t-t_s}{\tausyn} \right) \right]^2 } \right. \nonumber \\
                  &+ \left. \expect{ \left[ w_i \Theta(t-t_s) \exp \left(-\frac{t-t_s}{\tausyn} \right) \right]}^2  \right\} \nonumber \\
                  &= \lim_{T \to \infty} \nu_i T \left\{ \frac{1}{T} w_i^2 \int_0^T \exp \left( -2\frac{t}{\tausyn} \right) \, dt \right. \nonumber \\
                  &- \left. \frac{1}{T^2} \left[ \int_0^T \exp \left( -\frac{t}{\tausyn} \right) \, dt \right] \right\} \nonumber \\
                  &= \frac12 w_i^2 \nu_i \tausyn \quad ,
\end{align}
so the relative fluctuations of $\gsyn$ are of the order
\begin{equation}
    \frac{\sqrt{\var{\gsyn_i}}}{\expect{\gsyn_i}} = \sqrt{\frac{1}{2 \, \nu_i \tausyn}}
\end{equation}
and vanish in the limit of large firing rates.

This warrants an expansion of Eq.~(\ref{eq:ufromg}) in $\Delta \gsyn_i , \; \forall i$. Considering only the first-order term we obtain
\begin{equation}
    u(t) = \frac{\Iext + \gl\El + \sum_i \gsyn_i(t) \Erev_i}{\expect{\gtot}} \quad ,
    \label{eqn:fastu(t)}
\end{equation}
which renders $u$ simply a linear transformation of the synaptic noise current $\Jsyn = \sum_i \gsyn_i \Erev_i$.

\subsection{Appendix III:\\ Derivation of the equivalence to an OU process}

From Eq. (\ref{eqn:gsynODE}), we can find that the synaptic noise $\Jsyn$ obeys the first-order inhomogenous ODE
\begin{equation}
    \frac{d\Jsyn}{dt} = -\frac{\Jsyn}{\tausyn} + \sum_{\mathrm{syn} \, i} \sum_{\mathrm{spk} \, s} \Delta \Jsyn_i \delta(t-t_s) \; ,
    \label{eqn:IsynODE}
\end{equation}
where $\Delta \Jsyn_i = w_i \Erev_i$.
This equation is highly reminiscent of the ODE that defines the OU process
\begin{equation}
    dx(t) = \theta \left[ \mu - x(t) \right] dt + \sigma dW(t) \; .
\end{equation}
It is well-known that the PDF of the OU process
\begin{align}
    f(x,t|x_0) &= \sqrt{\frac{\theta}{\pi \sigma^2 (1-e^{-2\theta t})}} \nonumber \\
               &\times \exp \left\{ \frac{-\theta}{\sigma^2} \left[ \frac{(x - \mu + (\mu-x_0) e^{-\theta t})^2}{1-e^{-2\theta t}} \right] \right\}
    \label{eqn:PDFOU}
\end{align}
is the unique solution of the Fokker-Planck equation
\begin{equation}
    \frac{1}{\theta}\frac{\partial f(x, t)}{\partial t} = \frac{\partial}{\partial x} \left[ (x - \mu) f \right] + \frac{\sigma^2}{2\theta} \frac{\partial^2 f}{\partial x^2}
    \label{eqn:FPOU}
\end{equation}
with starting condition $x_0:=x(t=0)$.
In the following, we prove that, under certain assumptions, the distribution of the synaptic input $\Jsyn$ obeys the same Fokker-Planck equation.
To this end, we follow an approach similar to \cite{ricciardi1979ouprocess}.

Consider the PDF of the synaptic input $f(\Jsyn, t)$. We can use the Chapman-Kolmogorov equation to describe its evolution after a short time interval $\Delta t$ as an integral over all possible intermediate states $J'$:
\begin{equation}
    f(\Jsyn, t + \Delta t) = \int_{-\infty}^\infty f(\Jsyn, t + \Delta t | J', t) f(J', t) dJ'
    \label{eqn:chapmankolmogorov}
\end{equation}
For a small enough $\Delta t$, the probability of the occurrence of multiple spikes within $\Delta t$ can be neglected.
As incoming spikes are assumed to be generated by Poisson processes, the probability of a single spike occurring in $\Delta t$ is $\Delta t \sum_i \nu_i$.
By summing over the two possible histories of $\Jsyn$ within $\Delta t$ (either a single incoming spike or no spike at all), we can use Eq. (\ref{eqn:IsynODE}) to find
\begin{align}
    &f(\Jsyn, t + \Delta t | J') \nonumber \\
    &= \left[ 1 - \Delta t \sum_i \nu_i \right] \delta \left[ \Jsyn - J' \exp \left( {-\frac{\Delta t}{\tausyn}} \right) \right] \nonumber \\
    &+ \Delta t \sum_i \nu_i \delta \left[ \Jsyn \! - \left( J' + \Delta \Jsyn_i \right)\exp \! \left( {-\frac{\Delta t}{\tausyn}} \right) \right] ,
\end{align}
where $\nu_i$ represents the afferent firing frequency at the $i$th synapse.
Plugging this into Eq. (\ref{eqn:chapmankolmogorov}) and integrating over $J'$ yields
\begin{align}
    &f(\Jsyn, t + \Delta t) \nonumber \\
    &= \left( 1 - \Delta t \sum_i \nu_i \right) \exp \left( \frac{\Delta t}{\tausyn} \right) f \left[ \Jsyn \exp \left( \frac{\Delta t}{\tausyn} \right), t \right] \nonumber \\
    &+ \Delta t \sum_i \nu_i \exp \left( \frac{\Delta t}{\tausyn} \right) f \left[ \Jsyn \exp \left( \frac{\Delta t}{\tausyn} \right) - \Delta \Jsyn_i, t \right] .
\end{align}
We can now expand $f(x, t + \Delta t)$ up to first order in $\Delta t$
\begin{align}
    &f(\Jsyn, t + \Delta t) \approx f(\Jsyn, t) + \left. \frac{\partial f(\Jsyn, t + \Delta t)}{\partial \Delta t} \right|_{\Delta t = 0} \Delta t
\end{align}
and rearrange the terms to obtain
\begin{align}
    &\frac{f(\Jsyn, t + \Delta t) - f(\Jsyn, t)}{\Delta t} \nonumber \\
    &= \left. \frac{\partial f(\Jsyn, t + \Delta t)}{\partial \Delta t} \right|_{\Delta t = 0} \nonumber \\
    &= \left\{ -\sum_i \nu_i \exp \left( \frac{\Delta t}{\tausyn} \right) f \left[ \Jsyn \exp \left( \frac{\Delta t}{\tausyn} \right), t \right] \right. \nonumber \\
    &+ \left(1 - \Delta t \sum_i \nu_i \right) \frac{1}{\tausyn} \nonumber \\
    &\hspace{5mm} \times \Bigg\{ \exp \left( \frac{\Delta t}{\tausyn} \right) f \left[ \Jsyn \exp \left( \frac{\Delta t}{\tausyn} \right), t \right] \nonumber \\
    &\hspace{9mm} + \exp \left( 2\frac{\Delta t}{\tausyn} \right) \Jsyn \frac{\partial f \left[ \Jsyn \exp \left( \frac{\Delta t}{\tausyn} \right), t \right]}{\partial \Jsyn \exp \left( \frac{\Delta t}{\tausyn} \right)} \Bigg\} \nonumber \\
    &+ \sum_i \nu_i \exp \left( \frac{\Delta t}{\tausyn} \right) f \left[ \Jsyn \exp \left( \frac{\Delta t}{\tausyn} \right) - \Delta \Jsyn_i, t \right] \nonumber \\
    &+ (\dots) \Delta t \Bigg\}_{\Delta t = 0}
\end{align}
By taking the limit $\Delta t \rightarrow 0$, we obtain:
\begin{align}
    \frac{\partial f(\Jsyn, t)}{\partial t} &= \frac{1}{\tausyn} \frac{\partial}{\partial \Jsyn} \left[ \Jsyn f(\Jsyn, t) \right] \nonumber \\
                                            &+ \sum_i \nu_i \left[ f(\Jsyn - \Delta \Jsyn_i, t) - f(\Jsyn, t) \right] \quad .
\end{align}
In the limit of small synaptic weights (i.e., $\Delta \Jsyn_i \rightarrow 0$), we can expand the second term on the RHS up to the second order in $\Delta \Jsyn_i$.
This yields, after some rearrangement
\begin{align}
    &\frac{\partial f(\Jsyn, t)}{\partial t} \nonumber \\
    &= \frac{1}{\tausyn} \frac{\partial}{\partial \Jsyn} \left[ \left( \Jsyn - \sum_i \nu_i \Delta \Jsyn_i \tausyn \right) f(\Jsyn, t) \right] \nonumber \\
    &+ \frac{\sum_i \nu_i {\Delta \Jsyn_i}^2}{2} \frac{\partial^2 f(\Jsyn, t)}{\partial {\Jsyn}^2} \quad ,
    \label{eqn:FPIsyn}
\end{align}
which is the exact equivalent of the Fokker-Planck equation of the OU process (Eq. \ref{eqn:FPOU}).
Since $u(t)$ is only a linear transformation of $\Jsyn(t)$, it can also be approximated by an OU process in the limit of large input frequencies and small synaptic weights, with Eq.~(\ref{eqn:fastu(t)}) and Eq.~(\ref{eqn:FPIsyn}) giving the specific time constant, mean value and variance:
\begin{align}
    & \theta = \frac{1}{\tausyn} \quad , \\
    & \mu = \frac{\Iext + \gl\El + \sum_i \nu_i w_i \Erev_i \tausyn}{\expect{\gtot}} \label{eqn:OUmu} \quad , \\
    & \frac{\sigma^2}{2} = \frac{\sum_i \nu_i \left[ w_i \left( \Erev_i - \mu \right) \right]^2 \tausyn}{2\expect{\gtot}^2} \quad .
\end{align}

We conclude this section with two important notes.
Firstly, for the above methodology to be generally appliccable, we must be able to take the limit $\Delta \Jsyn_i \rightarrow 0$ for arbitrary first and second moments of $f(\Jsyn, t)$ without modifying them.
This is possible if at least one excitatory and one inhibitory input is present, which then give us two degrees of freedom with a proper choice of $\nu^\mathrm{exc} \to \infty$ and $\nu^\mathrm{inh} \to \infty$.
Secondly, all higher moments (3 and above) need to vanish in the abovementioned limit.
This has been shown to also be the case under the above conditions \cite{lansky1997sources}.

\subsection{Appendix IV:\\ Derivation of the activation function}

We can distinguish between two firing modes of the neuron.
The first mode can be classified as ``burst spiking'' and occurs when multiple spikes occur in rapid succession with an expected ISI of $\expect{\Delta t_k} = \tauref + \expect[s]{t_k^{s+1}-t_k^s} = \tauref + \taubk$, where $\taubk$ represents the average drift time from the reset to the threshold potential following the $k$th refractory period within a burst.
In this case, for each spike within a burst,
\begin{equation}
    \ueff(t_s) \geq \uthr \quad ,
\end{equation}
and also, for all but the last spike,
\begin{equation}
    \ueff(t_s + \tauref) \geq \uthr \quad .
\end{equation}
The second mode appears between such bursts, where the membrane potential evolves freely in the subthreshold regime.
If we define, just like in the abstract model, that the $k$th neuron is in the state $z_k=1$ for a duration $\tauon=\tauref$ following a spike, we can write
\begin{equation}
    p(z = 1) = \frac{\sum_n P_n \cdot n \cdot \tauref}{\sum_n P_n \cdot \left(n\tauref + \sum_{k=1}^{n-1} \taubk + T_n\right)} \quad ,
    \label{eqn:activationburstsum}
\end{equation}
where $P_n$ represents the distribution of burst lengths (conditioned on the existence of the first spike) and $T_n$ is the mean time interval between the end of a burst (i.e., the endpoint of its last refractory period) and the next spike.
The variables $P_n$, $T_n$ and $\taubk$ depend on all neuron and noise parameters, but for calculating the activation function (Fig.~2 in the main manuscript and Fig.~\ref{fig:actfctcomp} in the Appendix), we only vary $\bar u$.

We can now calculate both $P_n$ and $T_n$ iteratively.
The idea behind this approach is to propagate the membrane potential PDF from spike to spike within a burst and cut off the irrelevant parts for a particular burst length~$n$.
We denote the spike times within a burst of length~$n$ by $t_0, \dots, t_{n-1}$ and the endpoint of such a burst by $t_n := t_{n-1} + \tauref$.
For brevity, we also use $u_i := u(t_i)$.
Assuming a first spike at some time $t_0$ ($u_0 := u(t_0) = \uthr$), a ``burst'' of length $n=1$ requires a subthreshold free membrane potential after the first refractory period ($u_1 := u(t_0 + \tauref) < \uthr$).
This occurs with probability
\begin{align}
    P_1 :&= p(u_1 < \uthr | u_0 = \uthr)\nonumber \\
        & = \underbrace{\int_{-\infty}^{\uthr} du_1 p(u_1 | u_0 = \uthr)}_{\mathcal I_1} ,
    \label{eqn:P1}
\end{align}
where $p(u_{i+1} | u_{i}) := f(u, \tauref | u_i)$, which was defined in Eq.~(\ref{eqn:PDFOU}).
On average, the neuron then stays in the subthreshold regime for a period equal to the mean first passage time from $u_1$ to $\uthr$, so the mean duration of the time interval until the onset of the next burst can be expressed as
\begin{align}
    T_1 & = \int_{-\infty}^{\uthr} du_1 p(u_1 | u_0 = \uthr) \expect{T(\uthr, u_1)} \quad .
\end{align}
The first-passage time problem of the OU process has often been discussed \cite{thomas1975meanfpt}.
While no closed-form expression for the distribution of first-passage times $T(b,a) = \inf{t \geq 0: x(t) = b | x(0) = a}$ is known, its moments can be computed analytically \cite{ricciardi1988fptdensity}.
In particular, the mean first passage time reads
\begin{align}
    &\expect{T(b,a)} \nonumber \\
    &= \frac{\theta}{\sigma} \sqrt{\frac{\pi}{2}} \int_a^b \!\! dx \exp \! \left[ \frac{(x-\mu)^2}{2\sigma^2} \right] \! \left[ 1 \! + \! \mathrm{erf} \! \left( \frac{x-\mu}{\sqrt{2} \sigma} \right) \right] .
    \label{eqn:FPTOU}
\end{align}

A burst of $n=2$ spikes can only occur when the effective membrane potential lies above the spiking threshold ($u_1 \geq \uthr$) after the first refractory period and below after the second ($u_2 < \uthr$).
This makes $P_2$ and $T_2$ recursive functions of $P_1$:
\begin{align}
    P_2 &= p(u_2 < \uthr , u_1 \geq \uthr | u_0 = \uthr) \nonumber \\
        &= p(u_1 \geq \uthr | u_0 = \uthr) \; p(u_2 < \uthr | u_1 \geq \uthr , u_0 = \uthr) \nonumber \\
        &\stackrel{\mathclap{(\mathrm{Eq.} \ref{eqn:P1})}}{=} \;\;\; \underbrace{(1 - \mathcal I_1)}_{=1-P_1} \underbrace{\int_{\uthr}^\infty \!\!\! du_1 p(u_1 | u_1 \geq \uthr) \left[ \int_{-\infty}^{\uthr} \!\!\! du_2 p(u_2 | u_1) \right]}_{\mathcal I_2} \label{eqn:P2}\\
    T_2 &= \int_{\uthr}^\infty du_1 p(u_1 | u_1 \geq \uthr) \nonumber \\
        &\times \left[ \int_{-\infty}^{\uthr} du_2 p(u_2 | u_2 > \uthr, u_1) \expect{T(u_2, \uthr)} \right],
\end{align}
where $p(u_i | u_i \geq \uthr)$ is a shorthand notation for $p(u_i | u_i \geq \uthr, u_{i-1} \geq \uthr, \dots, u_1 \geq \uthr, u_0 = \uthr)$.
In particular, this represents a renormalization of the PDF of the effective membrane potential to values above the spiking threshold after $i$ refractory periods.

We can now continue this recursion up to an arbitrary burst length and write
\begin{align}
    P_n &= p(u_n < \uthr , u_{n-1} \geq \uthr, \dots, u_1 \geq \uthr | u_0 = \uthr) \nonumber \\
        &= p(u_1 \geq \uthr | u_0 = \uthr) \nonumber \\
        &\times p(u_n < \uthr , u_{n-1} \geq \uthr, \dots, u_2 \geq \uthr | u_1 \geq \uthr , u_0 = \uthr) \nonumber \\
        &\stackrel{\mathclap{(\mathrm{Eq.} \ref{eqn:P1})}}{=} \;\;\; (1-\mathcal I_1) \; p(u_2 \geq \uthr | u_1 \geq \uthr , u_0 = \uthr) \nonumber \\
        &\times p(u_n < \uthr , u_{n-1} \geq \uthr, \dots, u_3 \geq \uthr | \nonumber \\
        & \hspace{40mm} | u_2 \geq \uthr , u_1 \geq \uthr , u_0 = \uthr) \nonumber \\
        & \stackrel{\mathclap{(\mathrm{Eq.} \ref{eqn:P2})}}{=} \;\;\; (1-\mathcal I_1) (1-\mathcal I_2) \; p(u_3 \geq \uthr | u_2 \geq \uthr , u_1 \geq \uthr , u_0 = \uthr) \nonumber \\
        &\times p(u_n < \uthr , u_{n-1} \geq \uthr, \dots, u_4 \geq \uthr | \nonumber \\
        & \hspace{40mm} | u_3 \geq \uthr , \dots, u_1 \geq \uthr , u_0 = \uthr) \nonumber \\
        & = \prod_{i=1}^{n-1} (1-\mathcal I_i) \; p(u_n < \uthr | u_{n-1} \geq \uthr, \dots, u_1 \geq \uthr , u_0 = \uthr) \label{eqn:Pn} \\
        & = (1 - \sum_{i=1}^{n-1} P_i) \nonumber \\
        &\times \underbrace{\int_{\uthr}^\infty du_{n-1} p(u_{n-1} | u_{n-1} \geq \uthr) \left[ \int_{-\infty}^{\uthr} du_n p(u_n | u_{n-1}) \right]}_{\mathcal I_n} \label{eqn:fullrecursionP} \\
    T_n &= \int_{\uthr}^\infty du_{n-1} p(u_{n-1} | u_{n-1} \geq \uthr) \nonumber \\
        &\times \left[ \int_{-\infty}^{\uthr} du_n p(u_n | u_n < \uthr, u_{n-1}) \expect{T(u_n, \uthr)} \right]
    \label{eqn:fullrecursionT}
\end{align}
The transition from a product to a sum between Eq.~(\ref{eqn:Pn}) and Eq.~(\ref{eqn:fullrecursionP}) requires the identity
\begin{equation}
    \prod_{i=1}^{n-1} (1-\mathcal I_i) = 1 - \sum_{i=1}^{n-1} P_i \quad ,
\end{equation}
which can be easily shown by induction from $P_n = \mathcal I_n \prod_{i=1}^{n-1} (1-\mathcal I_i)$ (Eq.~\ref{eqn:Pn}) and $P_1 = \mathcal I_1$ (Eq.~\ref{eqn:P1}).
Since $\lim_{n \to \infty} P_n = 0$, one can stop the recursion at some small enough $P_n$.

What remains to be calculated is the average time-to threshold $\taubk$ within a burst that follows the $k$th refractory period.
Since we assume a HCS, we are looking at a regime in which $\taueff \ll \tausyn$.
Therefore, we can assume $\ueff$ to be approximately unchanged during the short time interval $\taubk$ (adiabatic approximation, see also \cite{moreno2004role}).
For a fixed $u_k$, the jump time can be easily calculated from Eq.~(\ref{eqn:reduceddudt}):
\begin{equation}
    \tau^{\mathrm b}_k(u_k) = \ln \left(\frac{\ureset-u_k}{\uthr - u_k}\right) \quad .
\end{equation}
The average jump time can then be obtained by integrating over all suprathreshold values of $u_k$, which in turn have probabilities that follow from integrating over all suprathreshold values of $u_{k-1}$ :
\begin{equation}
    \taubk = \int_\uthr^\infty du_k \ln \left(\frac{\ureset-u_k}{\uthr - u_k}\right) \int_\uthr^\infty du_{k-1} p(u_k | u_k > \uthr, u_{k-1}) \; .
    \label{eqn:taubk}
\end{equation}

With Eq.~(\ref{eqn:activationburstsum}), (\ref{eqn:fullrecursionP}), (\ref{eqn:fullrecursionT}), (\ref{eqn:taubk}) and (\ref{eqn:FPTOU}), one could now predict the activation function of an LIF unit in an extreme high-noise regime ($\taueff \to 0$).
We can, however, generalize our approach by taking the finite nature of the effective time constant into account.

If we go back to Eq.~(\ref{eqn:reduceddudt}) and leave $\taueff = C / \expect{\gtot}$ small but finite, we can still perform all the remaining approximations, but are required to modify Eq.~(\ref{eqn:fastu(t)}):
\begin{equation}
    \taueff \dot{u}(t) = \frac{\Iext + \gl\El}{\expect{\gtot}} + \frac{\Jsyn(t)}{\expect{\gtot}} - u(t) \quad .
\end{equation}
Together with Eq.~(\ref{eqn:IsynODE}), we now have a system of first-order ODEs which can be solved analytically by standard techniques (variation of constants).
The PSPs are then no longer a linear transformation of the exponentially shaped PSCs, but rather alpha-shaped (more precisely, a difference of exponentials):
\begin{equation}
  u_s(t) = \Theta(t-t_s) A \frac{\left( e^{-\frac{t-t_s}{\taueff}} - e^{-\frac{t-t_s}{\tausyn}} \right)}{\taueff - \tausyn} \quad ,
  \label{eqn:alphashape}
\end{equation}
with $A = \frac{w_i (\Erev_i - \expect{\ueff}) {\tausyn}_i}{\expect{\gtot}}$.
This shape causes a lower PSP peak than in the case of exponential PSPs, decreasing the overall width of the membrane potential distribution.
Intuitively speaking, this results in a horizontal shift and compression of the activation function.

More recently, analytical treatments of these phenomena have been proposed \cite{burkitt2006review}.
In these approaches, large membrane time constants (equivalent to a long $\taueff$) and small synaptic time constants are usually considered.
However, Eq.~(\ref{eqn:alphashape}) is symmetric in $\taueff$ and $\tausyn$, so the same argument applies to our case as well, but the two time constants need to be switched.
It is, for example, possible to correct the first passage time from the reset to the threshold potential by using an expansion in $\sqrt{\tau'/\tau}$ (with $\tau'$ and $\tau$ being the smaller and the larger of the two time constants, respectively) \cite{brunel1998firingfrequency}:
\begin{equation}
  \expect{T(\uthr, u)} = \tau \sqrt{\pi} \int_{\frac{u-\mu}{\sigma}}^{\frac{\thetaeff-\mu}{\sigma}} dx \exp(x^2)[\erf(x) + 1] \quad ,
  \label{eqn:newpassagetime}
\end{equation}
with $\mu$ and $\sigma^2$ the first two moments of the free membrane potential distribution and an effective threshold
\begin{equation}
    \thetaeff \approx \uthr - \zeta\left(\frac{1}{2}\right) \sqrt{\frac{\tau'}{2\tau}} \sigma \quad ,
\end{equation}
in which $\zeta$ denotes the Riemann zeta function.
In our particular case, the expansion is done in $\sqrt{\taueff/\tausyn}$, so $\tau'=\taueff$ and $\tau=\tausyn$.
With this approximation, we assume that $u$ converges from $\rho$ to $\ueff$ in negligible time after it is released from the refractory state.
Afterwards, its convergence to $\ueff$ is determined by Eq.~(\ref{eqn:reduceddudt}). 
Note how Eq.~(\ref{eqn:newpassagetime}) is equivalent to a change of the integration variable and limits in the original equation (Eq.~\ref{eqn:FPTOU}) for the first passage time.

\subsection{Appendix V:\\ Translation of synaptic weights}

A sufficient condition for a single neuron to sample from the correct conditional distribution is given by its activation function:
\begin{equation}
    p(z_k = 1 | z_{\setminus k}) = \sigma (v_k) \quad .
\end{equation}
The relationship between the abstract model and the LIF implementation is defined by the lateral dilation $\alpha$ and relative offset $\bar u_k^0$ of the LIF activation function:
\begin{equation}
    \label{eqn:activation_function_lif}
    p(z_k = 1 | z_{\setminus k}) = \sigma \left(\frac{\bar u_k - \bar u_k^0}{\alpha}\right) \quad .
\end{equation}
The parameters $\alpha$ and $\bar u_k^0$ can be determined by fitting Eq.~(\ref{eqn:activation_function_lif}) either to simulation results or to the theoretical prediction (Eq. \ref{eqn:activationburstsum}).
As a consequence, also synaptic weights need to be rescaled by the factor $\alpha$.
Additionally, the difference in PSP shapes needs to be taken into account.  

We choose a translation rule in which we set the LIF synaptic weights $w_{ij}$ such that the area under a PSP (Eq.~\ref{eqn:alphashape}) during the refractory state of the corresponding afferent neuron (i.e., for a duration $\tauref$) is equal to $ W_{ij} \, \tauref \, \alpha$:
\begin{align}
  W_{kj} &\tauref \alpha = \int_{0}^{\tauref} \frac{w_{kj} (\Erev_{kj} - \expect{\ueff}) \tausyn}{\expect{\gtot}} \nonumber \\
         &\times \frac{ \exp \left( -\frac{t-t_s}{\taueff} \right) - \exp \left(-\frac{t-t_s}{\tausyn} \right)}{\taueff - \tausyn} dt \nonumber \\
         &= \frac{w_{kj} \tausyn}{\expect{\gtot}} \frac{\left(\Erev_{kj} - \mu \right)} {\taueff - \tausyn} \nonumber \\
         &\times \left[\tausyn \left( e^{-\frac{\tauref}{\tausyn}} - 1 \right) - \taueff \left( e^{- \frac{\tauref}{\taueff}} - 1 \right) \right] \quad.
\end{align}
By setting $\tauref = \tausyn$, we obtain the mapping between the abstract and LIF synaptic weight domains:
\begin{align}
    W_{kj} &= \frac{1}{\alpha \Cm} \frac{w_{kj} \left(\Erev_{kj} - \mu\right)}{1 - \frac{\tausyn}{\taueff}} \nonumber \\
           &\times \left[\tausyn \left( e^{-1} - 1 \right) - \taueff \left( e^{-\frac{\tausyn}{\taueff}} - 1 \right) \right] \quad.
    \label{eqn:bm_to_nn}
\end{align}
Additionally, depressing short-term plasticity \cite{tso97} has been applied to attenuate the amplitudes of consecutively arriving alpha-shaped PSPs from a network neuron and emulate renewing synapses.
In particular, within the Tsodyks-Markram short-term plasticity model \cite{tso97}, the synaptic efficacy parameter and recovery time constant have been chosen as $U_\mathrm{SE}=1$ and $\tau_\mathrm{rec}=\tausyn$, respectively.

Analogously to the weights, the biases can be determined from the condition $\sigma(b_k) = \sigma(\left[\bar u_k^b - \bar u_k^0\right]/\alpha)$ in the absence of recurrent activity, i.e.\ $\Irec_k = 0$.

\subsection{Appendix VI:\\ Probabilistic model for demonstration of inference}

We define a joint model $p(\v y, \v z)$ over real-valued input nodes $\v y = (y_1, \dots, y_K)$ and binary latent variables $\v z  = (z_1, \dots, z_K)$ as sketched in Fig.~4A of the main manuscript.
The real-valued variables $y_k$ encode intensities of the input pixels.
The latent variables $z_k$ correspond to neurons in the network.
Our aim is to demonstrate that the network can sample from the posterior distribution $\cp{\v z}{\v y}$.

For this example, we have chosen a particularly simple likelihood function, namely a Gaussian emission model with variance $\sigma^2=1$ and mean values $\mu = \pm \frac 12$. 
From the graphical model we identify the structure of the joint distribution
\begin{align}
  p(\v y, \v z) &= p(\v z) \cdot \prod_{k=1}^{K} \cp{y_k}{z_k}
\end{align}
where $p(\v z)$ is a Boltzmann distribution and the likelihood is defined by
\begin{align}
    \cp{y_k}{z_k} &= \normal[\mu = 1/2]{y_k}^{z_k} \cdot \normal[\mu = - 1/2]{y_k}^{1-z_k} \quad ,
\end{align}
which is equivalent to
\begin{align}
    \log \cp{y_k}{z_k} &= z_k \cdot \left[ \log \normal[1/2]{y_k} - \log \normal[- 1/2]{y_k} \right] \nonumber \\
                       &+ \log \normal[- 1/2]{y_k} \nonumber \\
                       &= z_k \cdot y_k  + \log \normal[- 1/2]{y_k}\;\;,
\end{align}
using normal distributions
\begin{equation}
    \normal[\mu]{y} := \exp[- (y - \mu)^2 / 2] \, / \, \sqrt{2 \pi}
\end{equation}
with unit variance.
The posterior of this model reads:
\begin{align}
  \cp{\v z}{\v y} &= \frac{p(\v y, \v z)}{p(\v y)} \nonumber \\
                  &= \exp\left( \frac 12 \v z\T \v W \v z + \v z\T \, (\v b + \v y) \right) \Big / \mathrm{Norm} \quad ,
\end{align}
where the normalizing constant depends on the input $\v y$, but is independent of the network variables and thus defines a Boltzmann distribution over $\v z$ for any $\v y$.
In particular, we identify the abstract membrane potential
$ v_k = b_k + y_k + \sum_{j} W_{kj} \, z_j $
for sampling from $\cp{\v z}{\v y}$ by means of a spiking network.

\subsection{Appendix VII:\\ Simulation parameters}
\label{sec:parameters}

All simulations have been performed with the NEURON simulation package \cite{hines06neuron} and the PyNN API \cite{davison08pynn}, with a time step of $dt = \unit[0.01]{ms}$. 
For the LIF neuron, we have chosen the following parameters (compare with, e.g., \cite{naud08} for parameters fitted to experimental data):
\begin{table}[h]
    \centering
    \begin{tabular}{ccl}
        \hline
        $\Cm$ & \unit[0.1]{nF} & membrane capacitance \\
        $\gl$ & \unit[5]{nS} & leak conductance \\
        $\El$ & \unit[-65]{mV} & leak potential \\
        $\rho$ & \unit[-53]{mV} & reset potential \\
        $\Erev_\mathrm{exc}$ & \unit[0]{mV} & excitatory reversal potential \\
        $\Erev_\mathrm{inh}$ & \unit[-90]{mV} & inhibitory reversal potential \\
        $\vartheta$ & \unit[-52]{mV} & threshold voltage \\
        $\tausyn$ & \unit[10]{ms} & synaptic time constant \\
        $\tauref$ & \unit[10]{ms} & refractory time constant \\
        \hline
    \end{tabular}
    \caption{
        Neuron parameters used for the simulations in the main manuscript.
        }
    \label{table:neuron_params}
\end{table}

Synaptic noise was implemented as bombardment by inhibitory and excitatory Poisson stimuli with rates $\nu_\mathrm{inh} = \nu_\mathrm{exc} = \unit[5000]{Hz}$.
The excitatory synaptic weight for the noise stimuli was set to $w^\mathrm{noise}_\mathrm{exc} = \unit[0.0035]{\mu S}$.
The inhibitory weight $w^\mathrm{noise}_\mathrm{inh}$ was adjusted as to yield $p(z_k=1) \approx 0.5$ with no current stimulus present. 
For above parameters, this happens at an average free membrane potential of $\bar u = \unit[-55]{mV}$.
This determines $w^\mathrm{noise}_\mathrm{inh}$ according to
\begin{equation}
    \left | \frac{E^\mathrm{rev}_\mathrm{inh} - \bar u}{ E^\mathrm{rev}_\mathrm{exc} - \bar u} \right | = \frac{w^\mathrm{noise}_\mathrm{exc}}{w^\mathrm{noise}_\mathrm{inh}} \quad .
    \label{eqn:w_ratio}
\end{equation}

\subsection{Appendix VIII:\\ The activation function of LIF neurons in a spiking noisy environment (Fig.~\ref{fig:2} in main manuscript)}
\label{sec:fig2}

In order to sweep through the activation function, the external current $\Iext$ was varied.
However, in order to facilitate a comparison with the logistic activation function of the abstract model, we have represented $p(z=1)$ as a function of $\bar{u}$ instead.
The latter is equivalent to the mean $\mu$ of the corresponding Ornstein-Uhlenbeck process, with Eq.~(\ref{eqn:OUmu}) allowing a direct translation between $\Iext$ and $\bar{u}$.

The abscissa values in Fig.~\ref{fig:2}D in the main manuscript represent averages of the free membrane potential obtained from 10 simulation runs with a total duration of $T_\mathrm{sim} = \unit[100]{s}$ and firing threshold $\theta$ set to $E^\mathrm{rev}_\mathrm{exc} = \unit[0]{mV}$.
The deviations from the theoretical prediction (Eq.~\ref{eqn:activationburstsum}) are smaller than the size of the symbols, therefore no errorbars are shown.

The ordinate values and standard errors were calculated from the simulated spike train data according to
\begin{align}
    p(z = 1) &= \frac{1}{N}\sum_{i=1}^{N} p_i \label{eqn:spks_p_convert} \quad , \\
    s &= \sqrt{\frac{1}{N-1} \cdot \sum_{i=1}^{N} \left[ p_{i} - p(z = 1) \right]^2} \label{eqn:act_fct_errors}\quad ,
\end{align}
with $p_i = \frac{N^\mathrm{spk}_i \tauon}{T_\mathrm{sim}}$ being the fraction of time spent in $z = 1$ and $N^\mathrm{spk}_i$ representing the total number of spikes in the $i$th out of $N=10$ performed simulations.
Since the respective standard errors of the mean are smaller than the size of the symbols, no error bars are shown.

\subsection{Appendix IX:\\ Sampling via recurrent networks of LIF neurons (Fig.~\ref{fig:3} and \ref{fig:3large} in main manuscript)}
\label{sec:fig3}

The simulated network consists of $K = 5$ neurons with a synaptic weight matrix $\v W$ and a bias vector $\v b$ (both in the Boltzmann domain).
All entries were drawn from a beta distribution $\mathcal B(0.5, 0.5)$ and mapped linearly to the interval $[-0.6, 0.6]$.
More specifically, $b_k, W_{kj} \sim 1.2 \cdot \left[ \mathcal B(0.5, 0.5) - 0.5 \right]$.
The parameters and mapping of the beta distribution were chosen with the intent of generating diverse distributions, spanning multiple orders of magnitude.
The bias $b_k$, defined in the Boltzmann domain, determines the probability $\cp{z_k = 1}{\v z \wo k = \v 0}$ for neuron $k$.
In the LIF domain, the probability $\cp{z_k = 1}{\v z \wo k = \v 0} = 0.5$ corresponds to the mean free membrane potential $\bar u^0_k$.
Then, a nonzero bias can be described in the LIF domain as a linear shift from $\bar u^0_k$ to a mean membrane potential $\bar u^b_k$.
This yields the linear transformation
\begin{align}
    b_k = & (\bar u^b_k - \bar u^0_k) / \alpha \quad ,
    \label{eqn:biastrans}
\end{align}
where $\alpha$ represents the scaling factor between the two domains.
Both quantities $\bar u^0_k$ and $\alpha$ can be determined from the predicted activation function of a single LIF unit.
The first quantity constitutes the inflection point of the activation function (at $\cp{z_k = 1}{\v z \wo k = \v 0}$ = 0.5), the latter follows from the slope of the function.

By computing $\bar u^b_k$, we can map any bias $b_k$ of a single unit of the Boltzmann machine onto a yet unconnected LIF neuron.
In simulations, $\bar u^b_k$ was established by injecting a temporally constant external current $I^k_{\mathrm{ext}}$ according to
\begin{equation}
    \Iext_k \! = \! (\alpha b_k + \bar u^0_k) \expect{\gtot} - \gl\El - \! \sum_i \nu_i w_i^\mathrm{noise} \Erev_i\tausyn .
    \label{eqn:u_Iext}
\end{equation}

In order to achieve sampling network dynamics in the LIF domain faithful to those displayed by an equivalent Boltzmann machine, the Boltzmann weight matrix $\v W$ was translated into LIF network weights $w_{ij}$ according to Eq.~(\ref{eqn:bm_to_nn}).
Thus, superposing PSPs saturate the membrane potential, approximating the constant amplitude of a PSP in the abstract neuron model.

For Fig.~\ref{fig:3}B in the main manuscript, this setup of a random Boltzmann machine was simulated $N=10$ times with different random seeds for the Poisson background for a duration of $T_{\mathrm{sim}} = \unit[10]{s}$.
The red bars show the analytically computed target joint distribution $p_\mathrm{B}(\v z)$.
The blue bars depict the network distribution $p_\mathrm{N}(\v z)$, calculated from the firing activity of the simulated LIF network set up to match $p_\mathrm{B}(\v z)$.
The means and error bars have been calculated as in Eq.~(\ref{eqn:spks_p_convert}) and (\ref{eqn:act_fct_errors}), respectively. \\

The above simulations were repeated with a significantly longer duration in order to study systematic deviations due to the LIF implementation.
Fig.~\ref{fig:3}C in the main manuscript shows the distance between the target distribution
$p_\mathrm{B}(\v z)$
and its LIF network representation $p_\mathrm{N}(\v z)$ in form of the
Kullback-Leibler divergence
\begin{equation}
  \Dkl{p_\mathrm{N}}{p_\mathrm{B}} = \expect[p_\mathrm{N}(\v z)]{\log \, [ \,
p_\mathrm{N}(\v z) \, / \, p_\mathrm{B}(\v z) \, ]} \
  \label{eqn:dkl}
\end{equation}
This estimate has been taken for one set of parameters ($\v W$, $\v b$) for ten independent trials (thin lines) in an LIF network at integration times $T$: $0 \leq T \leq T_{\mathrm{sim}} = \unit[10^6] {ms}$. 
The red dashed line displays the averaged $\Dkl{p_\mathrm{N}}{p_\mathrm{B}}$ for the abstract network model with identical parameters ($\v W$, $\v b$).
The decrease of $\Dkl{p_\mathrm{N}}{p_\mathrm{B}}$ for longer integration times indicates the increasing precision of the sampling network over time. Eventually, the $\Dkl{p_\mathrm{N}}{p_\mathrm{B}}$ converges for the LIF network to a nonzero value, reflecting small systematic errors.
Fig.~\ref{fig:3}D in the main manuscript shows the distribution of $\Dkl{p_\mathrm{N}}{p_\mathrm{B}}$ values for 100 randomly drawn Boltzmann machines emulated by LIF networks, evaluated from a single run of $T_{\mathrm{sim}} = \unit[10^6] {ms}$ each.

Fig.~\ref{fig:3large} was generated in the same way as Fig.~\ref{fig:3}, except for the network size and weight distribution.

\subsection{Appendix X:\\ Demostration of probabilistic inference\\ (Fig.~\ref{fig:4} in main manuscript)}

We trained a fully visible, fully connected Boltzmann machine to store three hand-written digits (0, 3, 4) that were taken from the MNIST data set \cite{mnist98} and were scaled down to 12x12 pixels.
The pixel intensities of these patterns (ranging from 0 to 1) were linearly
scaled to an activation between $0.05$ and $0.95$, defining the target
statistics $\expect[T]{z_k}$ and $\expect[T]{z_k\, z_j}$ for the Boltzmann
machine. Then, external currents $I_k^{b}$ and synaptic weights $w_{kj}$ were optimized via the update rules
$\Delta I_k^{b} \propto  \expect[T]{z_k} - z_k$ and 
$\Delta w_{kj} \propto \expect[T]{z_k\, z_j} - z_k\, z_j$, 
with samples $\v z$ obtained through sampling from an LIF network set up with
synaptic noise and neuron parameters as described above.

The recurrent connections, defined by $w_{kj}$, induce synaptic currents $\Irec_k$ in addition to the noise currents $\Inoise_k$. 
Additionally, each neuron's mean effective membrane potential Eq.~(\ref{eqn:OUmu}) is shifted by external currents of the trained quantities $I_k^{b}$ as well as input currents $I_k^y$.
The latter encode observations $\v y$ which are defined in the context of the probabilistic model described in AVI.
Hence, the total received current of neuron $k$ in the network amounts to
\begin{equation}
  I_k = \Irec_k + \Inoise_k + I^b_k + I^y_k .
\end{equation}
The resulting samples $\v z^{(t)}$ displayed in subplots \ref{fig:4}C and \ref{fig:4}D were taken for $\unit[4000]{ms}$, after a burn-in time that ensured that the network
had converged to its equilibrium distribution.\\

\paragraph{Figure \ref{fig:4}C, main manuscript: Sampling from the prior.}

From the probabilistic model it follows that $\cp{\v z}{\v y = \v 0} = p(\v z)$.
This means that the LIF network will sample from the prior when $I^y_k = 0 , \; \forall k$.
The figure makes use of two projections of network states $\v z^{(t)}$ to illustrate the sampled distribution:

\begin{enumerate}

\item Star plot:
In order to illustrate the 12x12-dimensional network states
$\v z^{(t)}$, a two-dimensional linear projection in a star plot has been chosen 
(blue dots). The axes indicate the three basis vectors $\v B$, representing
pixel intensities $\expect[T]{z_k}$ of the digits (0, 3, 4),
\begin{equation}
    \expect[T]{z_k}^{034} = ( \v B^0 , \v B^3, \v B^4 ) \T
\end{equation}
with a total intensity normalization $|| \v B^i || = \sqrt{\sum_j |
B^i_j|^2} = 1$.

The network states $z^{(t)}$ acquired from the simulation are projected onto
this basis:
\begin{equation}
    \v z^{034}(t) = (\v B^0 \cdot z^{(t)}, \v B^3 \cdot z^{(t)}, \v B^4 \cdot z^{(t)})\T \quad .
\end{equation}
This three-dimensional vector is then projected onto a two-dimensional plane
in coordinates
\begin{equation}
\v z^{\mathrm{proj}}(t) =
\begin{pmatrix}
  \sin(\phi^0_\mathrm{B}) & \sin(\phi^3_\mathrm{B}) & \sin(\phi^4_\mathrm{B}) \\
  \cos(\phi^0_\mathrm{B}) & \cos(\phi^3_\mathrm{B}) & \cos(\phi^4_\mathrm{B})
 \end{pmatrix}
 \v z^{034}(t)
\end{equation}
with $(\phi^0_{\mathrm{B}}, \phi^3_{\mathrm{B}}, \phi^4_{\mathrm{B}}) = (0, \frac{2\, \pi}{3}, \frac{4\,\pi}{3})$
indicating the directions of the normalized basis vectors.

These linear projections $\v z^{\mathrm{proj}}(t)$ of network states $\v z^{(t)}$ are
used to illustrate similarity of states as distance in the two-dimensional
plane. In Fig.~\ref{fig:4}C in the main manuscript, a network evolution time of $\unit[4000]{ms}$ is shown, samples 
$\v z^{(t)}$ taken every $\unit[2]{ms}$, i.e.\ in total 2000 projected states are displayed.
We ensured in longer simulations that the total simulation runtime is sufficient to represent the
distribution under the mixing of the Markov chain.

The significant clustering of the network states around the directions of the arrows
indicates the proximity to the target states (0, 3, 4) for a majority
of time.

The transitions of the network states $\v z^{(t)}$ are depicted as a red
trajectory in the star plot. This trajectory connects 100 projected
network states within a time interval of $\unit[200] {ms}$, demonstrating the time evolution of
$\v z^{\mathrm{proj}}(t)$.

\item Snapshots:
The squared color maps on the time axis display the
averaged pixel intensity of the network in a time window $\unit[25] {ms}$:
\begin{equation}
    \bar z_k(t) = \frac{1}{\unit[25] {ms}} \int_{t-\unit[12.5] {ms}}^{t+ \unit[12.5] {ms}} z_k^{(t')} \, dt' .
\end{equation}
The equidistant time intervals between these snapshots were taken every $\unit[25]
{ms}$ within the time frame of the above mentioned red trajectory of $\unit[200] {ms}$.

\end{enumerate}

In summary, the network spends most of the time in distinct modes (digits) and only little time in ``blurred'' states that could not be clearly assigned to one digit. Furthermore, it spends an approximately equal amount of time in each mode, indicating that the prior $p(\v z)$ does not favor one of the three digits.\\

\begin{figure*}
    \centering
     \includegraphics[width=2.10\columnwidth]{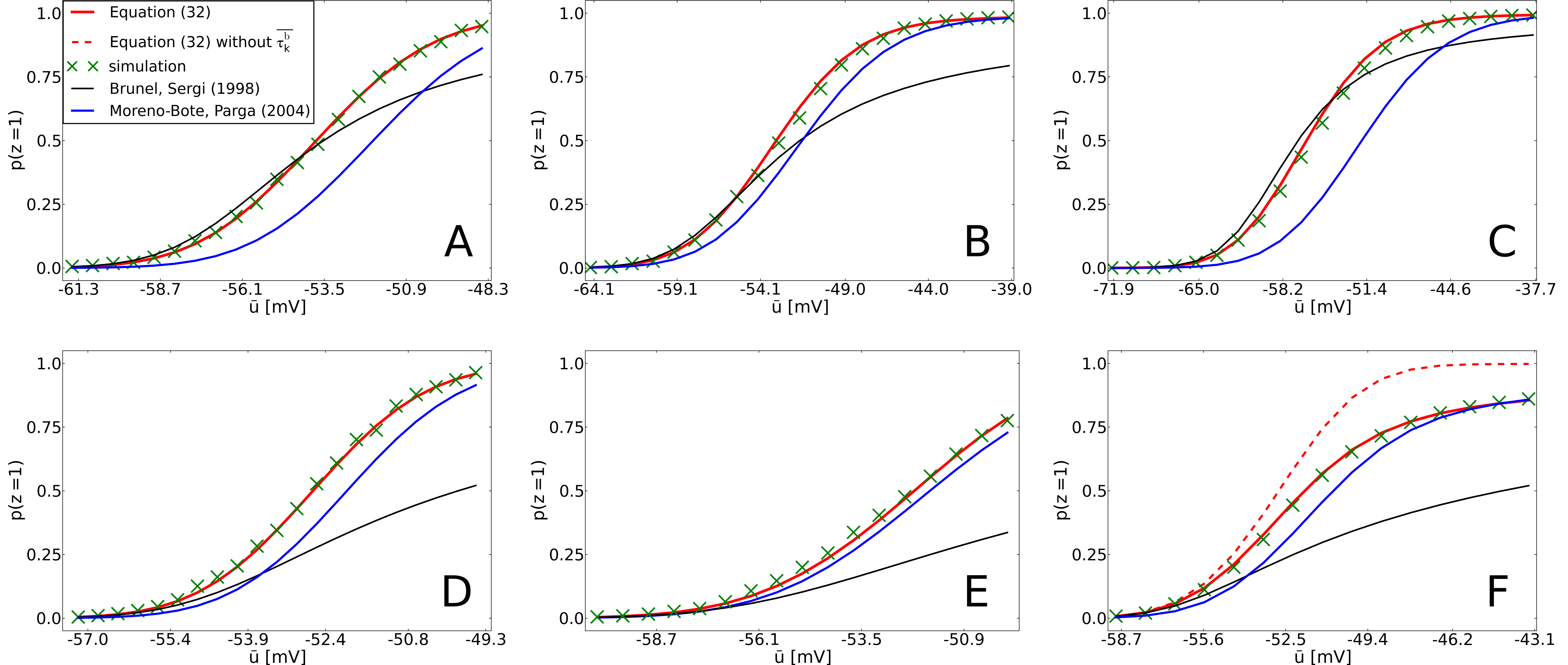}
    \caption{
        Comparison of our prediction of the activation function (Eq.~\ref{eqn:activationburstsum}) to simulation data, as well as to the predictions given by \cite{brunel1998firingfrequency} and \cite{moreno2004role}, for several parameter sets.
        (A) Standard parameter set as given in Tab.~\ref{table:neuron_params}.
        This figure shows the same curves as Fig.~\ref{fig:2}B in the main manuscript.
        (B) Same as A, but with quadrupled membrane capacitance $\Cm$ and one quarter of the leak conductance $\gl$, input rates $\nu_\mathrm{exc,inh}$ and weights $w_\mathrm{exc,inh}$.
        This parameter set is identical to the one used for the top left corner of Fig.~\ref{fig:paramsweep} and has the effect of slowing the membrane, i.e., increasing $\taueff$ by a factor of 16.
        (C) Same as A, but with a decreased synaptic time constant $\tausyn=\unit[3]{ms}$.
        The prediction from \cite{brunel1998firingfrequency} is improved, since the correlations in the pre- and post-refractory effective membrane potential are smaller in this scenario.
        (D) Same as A, but with an increased synaptic time constant $\tausyn=\unit[30]{ms}$.
        The prediction from \cite{brunel1998firingfrequency} deteriorates due to the longer-range membrane potential autocorrelation.
        Conversely, the prediction from \cite{moreno2004role} improves, since the refractory time becomes less important.
        (E) Same as A, but with a very short refractory time $\tauref=\unit[1]{ms}$.
        Here, we enter the parameter range where \cite{moreno2004role} provides good predictions.
        (F) Same as A, but with the input rates $\nu_\mathrm{exc,inh}$ and weights $w_\mathrm{exc,inh}$ decreased by a factor of 10, thereby slowing the membrane considerably (imperfect HCS).
        Additionally, we have chosen a large reset-to-threshold distance of $\uthr - \ureset = \unit[10]{mV}$.
        In this scenario, the $\taubk$-term in Eq.~(\ref{eqn:activationburstsum}) becomes dominant and the activation function departs from the logistic shape that it has in the HCS.
        }
    \label{fig:actfctcomp}
\end{figure*}

\paragraph{Figure \ref{fig:4}D, main manuscript: Sampling from the posterior.}

Incomplete and ambiguous input $\v y$ was provided to the network by setting
four of the inputs different from zero:  
$y_k \neq 0$ for $k \in \mathcal I$ with $\mathcal I$ denoting an index set and
$\vert{\mathcal I}\vert = 4$.
These inputs were chosen at the center of the image where digits 3 and 4 both have black pixels, while digit 0 is white.
More precisely, $\mathcal I = \{77, 78, 79, 80 \}$ when pixels are indexed
row-wise starting from the top-left corner.
In the LIF implementation, positive values $y_k$ correspond to positive currents
$I^y_k$, as specified in Eq.~(\ref{eqn:u_Iext}).
We set $I^y_k = \unit[0.831]{nA}$ for the four non-zero inputs,
inducing
an effective bias $b_k + y_k$ by injecting a total current $I_k^b + I_k^y$.
For the plot, the same projections of network states were used as in Fig.~\ref{fig:4}C in the main manuscript.

Under the incomplete input, the network spends only little time in the 0-mode or ``blurred'' states, while the equilibrium distribution exhibits two distict modes in the 3- and 4-directions. Thus the posterior reflects both the almost certain conclusion that ``the input is not a zero'' and the uncertainty that ``the input could either be a three or a four''.  
In particular, the network response is well-suited for further processing (e.g.\
by other cortical populations or in a technical application). For instance, the
network states could be integrated by a linear classifier to recognize the digit
class.

\begin{figure*}
    \centering
    \includegraphics[height=140pt]{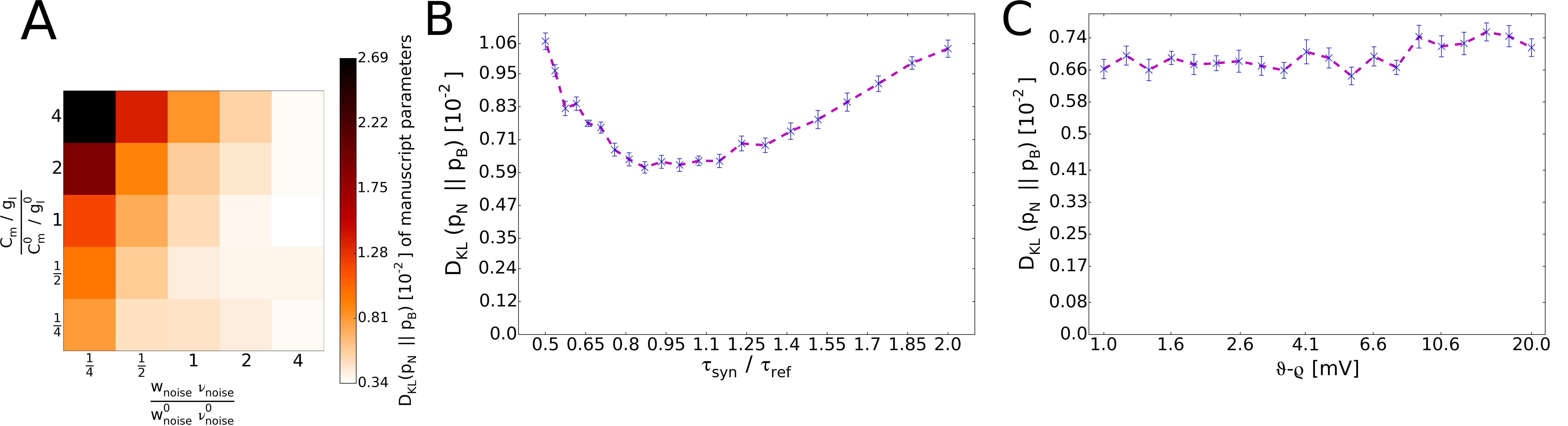}%[height=125pt]{Figure_7}
    \caption{
        Sampling with LIF neurons over a broad range of relevant model parameters.
        All plots depict the Kullback-Leibler divergence between the sampled distribution (with a certain set of parameters) and the target distribution, which is identical to the one used in Fig.~\ref{fig:3}B in the main manuscript.
        (A) Sweep over neuron size and leakage, as well as background input parameters.
            The axes represent multiplicative scaling values for four parameters: background (Poisson) synaptic weights $w_\mathrm{exc,inh}$ and firing rates $\nu_\mathrm{exc,inh}$ on the abscisa, neuron capacitance $\Cm$ and leak conductance $\gl$ on the ordinate.
            The parameter values used throughout the main manuscript (see also ``Appendix VII: Simulation parameters'') therefore have the coordinates $(1,1)$. 
            The network simulation runtimes were chosen as $T_{sim} = \unit[10^6]{ms}$.
            As expected, the large neuron / weak noise scenario (top left square) does not permit accurate sampling, as the activation function is no longer logistic (see also Fig.~\ref{fig:actfctcomp}F).
            In general, the plot shows that good sampling quality can be achieved for any neuron capacitance and leak as long as the background noise is strong enough (HCS).
        (B) Sweep over the ratio of the synaptic and refractory time constants.
            The best sampling performance is, indeed, achieved for $\tausyn \approx \tauref$, but the network still produces good approximations of the target distribution when the two time constants are not precisely identical.
            All $\DKL$ data points result from 20 simulations with runtimes of $T_{sim} = \unit[10^5]{ms}$ each.
            The error bars represent the standard error of the mean.
        (C) Effect of an increased distance from reset to threshold potential.
            An increase in $\uthr-\ureset$ causes a gradual decay of the sampling quality, since the membrane requires additional time to reach a suprathreshold $\ueff$ when the refractory period is over.
            This can, in principle, be accomodated by defining a larger time window $\tauon$ during which the neuron is considered to encode the state $z=1$.
            Nevertheless, in an HCS, the effective time constant can be low enough to render the threshold-to-reset distance irrelevant.
        }
    \label{fig:paramsweep}
\end{figure*}

\subsection{Appendix XI:\\ Prediction of the activation function for various parameter sets}

We have used Eq.~(\ref{eqn:activationburstsum}) for predicting the activation functions of LIF neurons in the HCS regime.
In this regime, the refractory time $\tauref$ and the synaptic time constant $\tausyn$ become the dominant time constants:
\begin{equation}
    \taueff \ll \tauref \approx \tausyn \quad .
\end{equation}
So far, we have compared our prediction to simulation data, as well as to two other predictions by \cite{brunel1998firingfrequency} and \cite{moreno2004role} in Fig.~\ref{fig:2}B of the main manuscript.
Here, we depart from these assumptions and show in Fig.~\ref{fig:actfctcomp} that our prediction holds for several different parameter sets.
In particular, the predictions from \cite{brunel1998firingfrequency} and \cite{moreno2004role} are only valid when either synaptic time constants (Fig.~\ref{fig:actfctcomp}C) or refractory times (Fig.~\ref{fig:actfctcomp}E) become shorter.

\subsection{Appendix XII:\\ Validity of the LIF sampling framework for various parameter sets}

For our various network simulations (Fig.~\ref{fig:3}-\ref{fig:4} in the main manuscript) we have used a set of biologically plausible parameters from \cite{naud08} (see also ``Appendix VII: Simulation parameters'').
In particular, we have used high input noise rates and weights in order to achieve the HCS, as well as a small distance between firing threshold and reset, in order to reduce $\taubk$, during which a neuron falsely encodes the state $z=0$.
Furthermore, we have assumed $\tauref=\tausyn$.
Fig.~\ref{fig:paramsweep} shows that these are by no means strict constraints.

\bibliography{refs}

\end{document}